\documentclass[11pt]{article}

\usepackage{amsmath,amsfonts,amsthm,amssymb,mathtools}
\usepackage{algorithm}
\usepackage{algpseudocode}
\usepackage{comment}
\usepackage{graphicx} % Required for inserting images
\usepackage[margin=1in]{geometry}
\usepackage{microtype}
\usepackage{xcolor}
\usepackage{enumitem}
\usepackage[colorlinks,citecolor=gray]{hyperref} % jack: do we want hidelinks? %I prefer colored links. % jack: me too, cool
\usepackage{cleveref}
\usepackage{xspace}

\crefname{appendix}{Appendix}{Appendices}

\newif\ifdraft
\drafttrue

\definecolor{draftred}{RGB}{155,30,30}
\definecolor{draftgray}{RGB}{245,245,245}

\ifdraft
  \newcommand{\draftnote}[1]{%
    \par\smallskip\noindent
    \fcolorbox{draftred}{draftgray}{%
      \parbox{0.94\linewidth}{\small\color{draftred}\textbf{Draft note.} #1}}
    \par\smallskip}
  \newcommand{\proofplaceholder}[1]{%
    \begin{proof}[Proof placeholder]
    \color{draftred}#1
    \end{proof}}
\else
  \newcommand{\draftnote}[1]{}
  \newcommand{\proofplaceholder}[1]{}
\fi

\newcommand{\norm}[1]{\|#1\|}

\newtheorem{theorem}{Theorem}
\newtheorem{lemma}{Lemma}
\newtheorem{proposition}{Proposition}
\newtheorem{corollary}{Corollary}

\theoremstyle{definition}
\newtheorem{definition}{Definition}

\newtheorem{question}{Question}

\theoremstyle{remark}

\crefname{theorem}{Theorem}{Theorems}
\crefname{lemma}{Lemma}{Lemmas}
\crefname{proposition}{Proposition}{Propositions}
\crefname{corollary}{Corollary}{Corollaries}
\crefname{conjecture}{Conjecture}{Conjectures}
\crefname{claim}{Claim}{Claims}
\crefname{definition}{Definition}{Definitions}
\crefname{problem}{Problem}{Problems}
\crefname{question}{Question}{Questions}
\crefname{remark}{Remark}{Remarks}

\DeclareMathOperator{\tr}{tr}

\DeclareMathOperator{\poly}{poly}
\newcommand{\R}{\mathbb{R}}

\newcommand{\eps}{\varepsilon}

\newcommand{\aspect}{\Delta}

\newcommand{\rank}{\operatorname{rank}}
 
  \usepackage{nth}
  \usepackage{intcalc}

  \newcommand{\cAAAI}[1]{AAAI\ Conference\ on\ Artificial (AAAI)}

\newenvironment{reptheorem}[1]{%
  \begingroup
  \renewcommand{\thetheorem}{\ref{#1}}%
  \renewcommand{\theHtheorem}{rep.#1}% unique hyperref anchor, so restatements do not steal the next theorem's link target
  \begin{theorem}\addtocounter{theorem}{-1}% restated theorems do not advance the counter
}{%
  \end{theorem}
  \endgroup
}

\newenvironment{replemma}[1]{%
  \begingroup
  \renewcommand{\thelemma}{\ref{#1}}%
  \renewcommand{\theHlemma}{rep.#1}% unique hyperref anchor, so restatements do not steal the next lemma's link target
  \begin{lemma}\addtocounter{lemma}{-1}% restated lemmas do not advance the counter
}{%
  \end{lemma}
  \endgroup
}

\title{Balanced Voronoi Partitions, Self-Approaching Sequences, and Applications to Efficiently Navigable Graphs}
\author{
Jonathan Conroy\thanks{Dartmouth College. \texttt{jonathan.conroy.gr@dartmouth.edu}}
\and
Mohammadreza Daneshvaramoli\thanks{University of Massachusetts Amherst. \texttt{mdaneshvaram@cs.umass.edu}, \texttt{cmusco@cs.umass.edu}}
\and
Cameron Musco\footnotemark[2]
\and
Christopher Musco\thanks{New York University. \texttt{cmusco@nyu.edu}}
\and
Jack Spalding-Jamieson\thanks{Independent. \texttt{jacksj@uwaterloo.ca}}
}
\date{}

\begin{document}

\maketitle

\begin{abstract}
    We study the following geometric question: given a set $X$ of points in $d$-dimensional Euclidean space, is it always possible to pick two points {$a,b \in X$} such that their induced Voronoi partition is approximately balanced? I.e., letting $X_a,X_b \subseteq X$ be the subsets of points closer to {$a$ and $b$} respectively, can we pick {$a$ and $b$} such that $\min(|X_a|,|X_b|) \ge \alpha \cdot |X|$ for some $\alpha > 0$?

Our main result shows that balance  $\alpha = \Omega(1/d)$ is always achievable, which is tight up to a constant factor.
    The proof is via a reduction to another natural question about \emph{self-approaching sequences} (SASs): what is the longest sequence of points in $d$-dimensional Euclidean space such that each successive point gets closer to all other points? We show that if there is no balanced Voronoi partition with balance $\alpha$, then one can find an SAS of length $\Omega(1/\alpha)$. We then prove that in $ d$-dimensional Euclidean space, any SAS has length at most $O(d)$, implying our balanced partitioning result. We discuss extensions beyond Euclidean space: surprisingly, for the $\ell_p$ norm for any constant $p \neq 2$,  we show that {one} cannot achieve balance better than $\exp(-\Omega(d))$. However, we do prove a $1/\poly(d)$ bound for point sets with bounded \emph{aspect ratio}. 

    While we seem to be the first to study the exact problems above formally, both are closely related to work on \emph{navigable search graphs}, which admit efficient routing based on greedy search, and form the backbone of many modern approximate nearest neighbor search systems. Results on navigable graphs in low-dimensional Euclidean space dating back to the 1990s can be used to give a bound of $\alpha = 1/2^{O(d)}$ for the balanced Voronoi partitioning problem. Our main result exponentially improves this bound, to $\Omega(1/d)$. We in turn use our bound to give the first construction of navigable graphs under a natural variant of classic greedy routing, with routing time $O(d \log |X|)$. All existing constructions require $\exp(d)$ routing time in the worst case.
\end{abstract}

\section{Introduction}\label{sec:introduction}

\begin{figure}[!ht]
\vspace{8pt}% extra room below Question 2
\centering
\begin{tabular}{@{}c@{\hspace{2.5em}}c@{\hspace{2.5em}}c@{}}
\includegraphics[page=1]{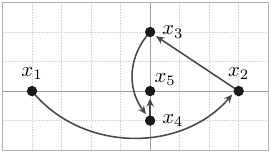} &
\includegraphics[page=2]{graphics/fig-sas-voronoi.pdf} &
\includegraphics[page=3]{graphics/fig-sas-voronoi.pdf} \\[2pt]
\small (a) An SAS in $\R^2$ &
\small (b) $a = x_1$, $b = x_3$ &
\small (c) $a = x_2$, $b = x_4$
\end{tabular}
\caption{(a) A self-approaching sequence $(x_1,\ldots,x_5)$ in $\R^2$. For every base point $x_k$, the other points get monotonically closer to $x_k$ as the sequence proceeds. (b)--(c) The Voronoi partitions of this SAS induced by two different pairs $a,b$. SASs give a hard case for Voronoi partitioning: all partitions are completely unbalanced, with one side having just a single point on it.}
\label{fig:sas-voronoi}
\end{figure}

In this work, we study two related geometric problems. The first concerns the idea of \emph{balanced Voronoi partitions}:

\begin{definition}[Balanced Voronoi Partition]\label{def:balancedPartition}
Let $X = \{x_1,\ldots,x_n\}$ be a set of $n \ge 2$ points in a metric space with distance function $D(\cdot, \cdot)$. $X$ admits an $\alpha$-balanced Voronoi partition if there exist distinct points $a,b \in X$ such that, letting $S(a,b)$ be the set of points in $X$ closer to $a$ and $S(b,a)$ be the set of points closer to $b$ (with any ties broken by ID), % = \{x \in X: D(x, x_1) \le D(x,x_2)\}$ and $S(x_2,x_1) = \{x \in X: D(x, x_2) \le D(x,x_1)\}$ be the sets of points closer to $x_1$ and $x_2$ respectively\footnote{$S(x_1,x_2)$ and $S(x_2,x_1)$ may not actually partition $X$: both contain points equidistant to $x_1$ and $x_2$. We can ensure a true partition if we assume that all pairwise distances are distinct, or break ties using an arbitrary ordering on the points in $X$. All our results on $\alpha$-balanced partitions will extend to this setting.},  
\begin{align*}
    \min\{|S(a,b)|,|S(b,a)| \} \ge \alpha \cdot n.
\end{align*}
\end{definition}
%See \Cref{fig:sas-voronoi}(b)-(c) for an illustration of .
Our main question is:
\begin{question}\label{question:q1}
    What is the largest $\alpha$ such that every set of points in $d$-dimensional Euclidean space has an $\alpha$-balanced Voronoi partition?
\end{question}
When $d = 1$, the answer is clear: assuming that $n$ is even, we can take $a$ and $b$ to be the two values adjacent to the median, and always achieve balance $\alpha = 1/2$. If $n$ is odd, we can achieve balance $\alpha = \frac{n-1}{2n} \approx 1/2$. However, in higher-dimensional Euclidean space, such an even split is generally not possible: there are worst-case point sets with no $\alpha$-balanced Voronoi partition for any $\alpha > 1/(2d)$. Indeed, a natural class of hard point sets are \emph{self-approaching sequences}:

\begin{definition}[Self-Approaching Sequence (SAS)]\label{def:sas}
Consider any metric space $(\mathcal X,D)$. 
    A sequence of distinct points $(x_1, \ldots, x_\ell) \in \mathcal X$ is a \emph{self-approaching sequence (SAS)} if for every $k$ and $i < j$ with $i \neq k$,
\[D(x_k,x_j) \le D(x_k,x_i).\]
In other words, for any choice of ``base point'' $x_k$, the sequence of points $(x_1, \ldots, x_\ell) \setminus \{x_k\}$ monotonically gets closer to $x_k$. See \Cref{fig:sas-voronoi}(a) for an example. %We say that $(x_1,\ldots,x_\ell)$ is a \emph{weak SAS} if for every $k$ and $i < j$ with $i \neq k$,
%$D(x_k,x_j) \le D(x_k,x_i).$ Any SAS is a weak SAS, but not vice-versa.
%Let \EMPH{\(L_d\)} denote the maximum possible length of a SAS in \(\R^d\).
\end{definition}

It is easy to see that in $d$-dimensional Euclidean space, there exist SASs of length $\Omega(d)$. E.g., consider a sequence of (rescaled) standard basis vectors $x_1 = r_1 \cdot e_1, x_2 = r_2 \cdot e_2, \ldots, x_d = r_d \cdot e_d$, for any scalings $r_1 > r_2 > \ldots > r_d > 0$. For any $k \neq j$, we have $\norm{x_k-x_j}_2^2 = r_k^2 + r_j^2$, and thus $\norm{x_k-x_j}_2^2 < \norm{x_k-x_i}_2^2$ whenever $i < j$ with $i \neq k$. In fact, one can check that the sequence $e_1, -\frac{1}{2} e_1, \frac{1}{4} e_2, -\frac{1}{8} e_2,\ldots$ gives an SAS of length $2d$.

Further, such a sequence gives the mentioned $\alpha \le \frac{1}{2d}$ upper bound for \Cref{question:q1}: the Voronoi partition for any chosen pair $x_i,x_j$ with $i < j$ will have just $1$ point closest to $x_i$ ($x_i$ itself) and all $2d-1$ other points closest to $x_j$. I.e., self-approaching sequences seem to give a worst-case for balanced Voronoi partitioning. See \Cref{fig:sas-voronoi}. Given this, we ask:
\begin{question}\label{question:q2}
    What is the maximum length of {an SAS} in $d$-dimensional Euclidean space?
\end{question}

\subsection{Our Contributions}

Our main contribution is to give bounds for \Cref{question:q1,question:q2} that are linear in the dimension $d$, and thus tight up to constant factors.
We first prove for \Cref{question:q2}:
\begin{theorem}[Linear Euclidean SAS Bound]\label{thm:Ld}
Let $L_d$ denote the maximum possible length of an SAS in $\R^d$ under the Euclidean distance. For every $d$, $L_d \le {8d}.$
\end{theorem}
As discussed, it is easy to see that {$L_d \ge 2d$}, so \Cref{thm:Ld} is tight up to a constant factor. It is also clear that $1/L_d$ gives an upper bound on the best achievable Voronoi partition balance $\alpha$ for general $d$-dimensional Euclidean point sets: given an SAS of length $L_d$, the Voronoi partition induced by any two points in the sequence has just one point assigned to the earlier point (itself) and all other points assigned to the later point (assuming we give later points in the sequence smaller IDs, so any distance ties are broken in favor of them). %. Then, in, every point other than the earlier one is assigned to the later one. 

We next  prove a converse statement (up to constants) for general metric spaces. %For technical reasons, this result holds for \emph{generic} point sets -- i.e., $X$ so that for any $a,b,c \in X$ with $b \neq c$, $D(a,b) \neq D(a,c)$. When $D$ is continuous (e.g., for Euclidean distance), it easily extends to arbitrary point sets.

\begin{theorem}[Balanced Partitioning to SAS Reduction]\label{thm:reduction}
Consider any metric space $(\mathcal X,D)$ and any set of points $X \subseteq \mathcal X$. Let $L_X$ denote the length of the longest SAS consisting of points in $X$. Then $X$ admits an $\alpha$-balanced Voronoi partition for $\alpha \ge 1/(4L_X)$.
\end{theorem}

That is, an upper bound on the maximum length of an SAS implies a lower bound on the worst-case Voronoi partition balance. To prove this, we show that, if $X$ does not have a sufficiently balanced Voronoi partition, then it must contain a subset of points that form a long SAS. 
%In \Cref{sec:reduction}, we also prove a variant of \Cref{thm:reduction}, \Cref{thm:reductionTie}, in which ties are broken according to a fixed ordering of the points, so that the two sides of the Voronoi partition truly partition $X$. In this variant, $L_{(\mathcal X,D)}$ is replaced by the length of the longest weak SAS. This is why we bound the lengths of weak SASs in \Cref{thm:Ld} and below, and the variant is what we use in our application to navigable graphs in \Cref{sec:navigation}.
Combining \Cref{thm:reduction} and \Cref{thm:Ld} (noting that $L_X \le L_d$ for any $X \subseteq \R^d$) gives our main result for balanced Voronoi partitions in Euclidean space:

\begin{theorem}[Euclidean Balanced Partitioning Bound]\label{thm:main}
Let $X$ be any set of points in $\R^d$ under the Euclidean distance. Then $X$ admits an $\alpha$-balanced Voronoi partition for $\alpha \ge {\frac{1}{32d}}$.
\end{theorem}

As we show in \Cref{app:exponential}, prior work in computational geometry on \emph{navigable graphs} in low-dimensional space \cite{AryaMount:1993} can be used to prove a bound of $\alpha = 1/2^{O(d)}$, and in turn $L_d = 2^{O(d)}$. \Cref{thm:main} exponentially improves this baseline, and as discussed, is tight up to constants.
%Further, it is tight up to a constant factor: as discussed, there are point sets in $\R^d$ with no $\alpha$-balanced Voronoi partition for any $\alpha > 1/d$.} %We conjecture that a tight bound of $\alpha = \Theta(1/d)$ in fact holds. % -- we leave proving this open.

\subsubsection{Extensions to Other {Metrics}}\label{sec:extensionsIntro}

It is natural to consider extensions of \Cref{thm:Ld} (and in turn \Cref{thm:main}) to other metric spaces. In particular, we study $\mathbb{R}^d$ under the $\ell_p$ norm for $p \neq 2$. Perhaps surprisingly, we show that in this case, \emph{no analog to \Cref{thm:Ld} holds}: there exist SASs with length exponential in $d$.

\begin{theorem}[Exponential SAS Lower Bound for $\ell_p$ Spaces]
\label{thm:negative-non-euclidean-p}
For every fixed $0<p\le \infty$ with $p\ne2$, there exists an SAS of length $\exp(\Omega_p(d))$ in $\R^d$ under the  $\ell_p$ norm.
\end{theorem}
To get some intuition behind \Cref{thm:negative-non-euclidean-p}, consider the case of $p = \infty$. Here, we can use a similar construction to the length-$d$ Euclidean SAS: take the $2^d$ vertices of the hypercube $\{-1,1\}^d$,  scaled by unique values $r_1 > r_2 >\ldots > r_{2^d} > 0$. The $\ell_\infty$ distance between any pair of points $x_i \neq x_j$  is exactly $r_i + r_j$. Thus, $\norm{x_k-x_j}_\infty < \norm{x_k-x_i}_\infty$ whenever $i < j$ with $i \neq k$, and we have an SAS. 

Alternatively, consider $p = 1$. Each of our vectors will correspond to one of $N=\exp(\Omega(d))$ subsets $X_1,\ldots,X_N\subseteq[d]$, each of size $m = \Theta(d)$ with $|X_i\cap X_j|\le m/4$ for all $i\ne j$. It is easy to construct such a family of sets -- e.g., random sets suffice with high probability. Set $r_i=3^{-i}$ and $x_i= r_i \cdot \mathbf 1_{X_i}$, where $\mathbf 1_{X_i} \in \{0,1\}^d$ is the indicator vector for $X_i$. For $k<j$, we can check:
\begin{align*}
 \norm{x_k-x_j}_1
 &=m \cdot r_k+ m \cdot r_j - 2|X_k\cap X_j| \cdot r_j\\
 &\in[m(r_k+r_j/2),m(r_k+r_j)]. % \in [9/8 a_j + 5/4 a_j] \le a_{i+1}.
\end{align*}
Due to the geometrically decreasing scalings of the points, the above bound ensures that for $k < i <  j$ the SAS condition (\Cref{def:sas}) holds for base point $k$: $\norm{x_k - x_i}_1 > \norm{x_k - x_j}_1$ since $r_i \ge 3r_j > 2r_j$. It is not hard to check the remaining cases (i.e., when $i < k < j$ and $i < j < k$), to show that the SAS condition holds in general. In \Cref{sec:lp-extensions} we use a similar construction to extend the lower bound to all constant $p \neq 2$, giving \Cref{thm:negative-non-euclidean-p}.

%Together with $a_{i+1}=a_i/4$, these bounds imply that the distance to any fixed base point strictly decreases along the remaining sequence. Thus $(x_1,\ldots,x_N)$ is an SAS; see \Cref{thm:negative-small-p} for the extension to $0<p\le1$.

\medskip

\noindent\textbf{Positive Results via Bounded Aspect Ratio.}
Despite the exponential lower bound of \Cref{thm:negative-non-euclidean-p}, we do show that SASs in $\ell_p$ spaces for fixed $1 \le p < \infty$ have length bounded by a polynomial factor in $d$ and a logarithmic factor in the \emph{aspect ratio} 
\begin{align}\label{eq:aspect}
\aspect(X):=\frac{\max_{x\ne y\in X}D(x,y)}{\min_{x\ne y\in X}D(x,y)}.
\end{align}
 The aspect ratio is a common parameter that arises e.g., in work on fast algorithms for nearest neighbor search and other geometric problems \cite{IM98,KrauthgamerLee:2004}. For SASs in $\R^d$ under any $\ell_p$ norm, letting $O_p(\cdot)$ and $\Omega_p(\cdot)$ hide constants depending on $p$, we show:
\begin{theorem}[Bounded Aspect Ratio SAS Bound -- $\ell_p$ Spaces]
\label{thm:bounded-aspect-intro}
For every fixed $1\le p<\infty$, let $q_p=\frac{2p+2}{2p-1}\le4$. Every SAS $X\subseteq\R^d$  under the $\ell_p$ norm with $\Delta(X) < b$ has length {$O_p\!\left(d^{3q_p/2}\lceil \log(b) \rceil \right)$}.
In turn, by \Cref{thm:reduction}, every $X\subseteq\R^d$ with $\Delta(X) < b$ admits an $\alpha$-balanced Voronoi partition under the $\ell_p$ norm with {$
 \alpha=\Omega_p\!\left(\frac{1}{d^{3q_p/2}\lceil\log(b)\rceil}\right)$}.
\end{theorem}

\Cref{thm:bounded-aspect-intro} follows from a more general upper bound on SASs with bounded aspect ratios that holds in any metric where one can show an upper bound on the size of a \emph{nearly equilateral set} -- i.e., a set of points with all pairwise distances nearly equal \cite{alon2003equilateral,swanepoel2004equilateral}. In fact, this general bound can also be used to prove a version of \Cref{thm:Ld} with a weaker $\poly(d)$ bound, by arguing that a general SAS must be comprised of $O(d)$ SASs, each with aspect ratio $O(d)$ (see \Cref{sec:euclidean-poly}).
\begin{theorem}[Bounded Aspect Ratio SAS Bound -- General Metrics]
\label{thm:bounded-aspect-general-intro}
Consider any metric space $(\mathcal X,D)$ and suppose that, for some $\eps>0$, every set of points in this space with aspect ratio $\le 1+\eps$ has size at most $N_\eps$. For any $b>1$, every SAS $X$ in $(\mathcal X,D)$ with $\Delta(X) < b$ has length at most
\[
 N_\eps\cdot\left\lceil\log_{1+\eps}b\right\rceil.
\]
\end{theorem}
\Cref{thm:bounded-aspect-intro} follows from applying \Cref{thm:bounded-aspect-general-intro} with $\eps = O(1/d^{q_p/2})$ -- existing bounds on exactly equilateral sets in $\ell_p$ spaces \cite{alon2003equilateral} can be extended to show $N_\eps = O(d^{q_p})$ in this case. 

We remark that our hard instance for the $\ell_1$ norm that establishes the exponential SAS length lower bound of \Cref{thm:negative-non-euclidean-p} has doubly exponential aspect ratio -- as is necessary given \Cref{thm:bounded-aspect-intro,thm:bounded-aspect-general-intro}. We also remark that our hard instance for $\ell_\infty$ is derived from an exponentially sized equilateral set, highlighting the importance of being able to bound $N_\eps$ in order to bound SAS length.

\subsubsection{Applications to Navigable Search Graphs}\label{sec:navigationIntro}

While balanced Voronoi partitions and self-approaching sequences are natural geometric objects in their own right, our work on both is motivated by their connections to \emph{navigable search graphs} \cite{AryaMount:1993,KrauthgamerLee:2004,diwan2024navigable,khanna2026sparse,conway2026efficiently,avi2026almost}. Such graphs form the basis of many current state-of-the-art methods for approximate nearest neighbor search, including the popular HNSW, DiskANN, and NSG algorithms \cite{SubramanyaEtAl:2019,FuXiangWangCai:2019,MalkovYashunin:2020}.

In the standard setup, one constructs a directed graph $G$ with vertices corresponding to each point in a dataset $X$. Greedy search for a query $q$ starts at some node and repeatedly moves to the out-neighbor of the current node that is closest to $q$, terminating if no neighbor improves on the current distance. $G$ is said to be  \emph{navigable} if given any \emph{in-dataset} query $q \in X$, greedy search terminates at $q$ when initialized at any starting point. 

Diwan et al.\ \cite{diwan2024navigable} show that, for any set of $n$ points in any metric space, there exists a navigable graph with average degree $O(\sqrt{n \log n})$. On the {other hand}, even for random point sets in $O(\log n)$ dimensional Euclidean space, this bound is nearly tight: navigability requires average degree $\Omega(n^{1/2-\delta})$ for any fixed $\delta > 0$. Thus, the total expected runtime required to perform greedy routing, which is lower bounded by the number of distance comparisons needed, and hence the degree of any vertex on the greedy path,  is $\Omega(n^{1/2-\delta})$. Daliri et al.~\cite{DaliriEtAl:2026} show that this bound holds even if one picks a best-case starting vertex.

A key question, especially given the success of greedy graph-based search in practical applications, is if stronger theoretical bounds can be shown for variants of the above theoretical model. E.g.,~\cite{DaliriEtAl:2026} demonstrate that strong sublinear  running time bounds can be given for approximate nearest neighbor search, if the search graph $G$ is augmented with Steiner points. Other works {give strong theoretical} guarantees, but with exponential dependence on the dimension (or doubling dimension) of the data set \cite{AryaMount:1993,Clarkson:1994,IndykXu:2023,lu2025proximity}. In fact, as discussed, we show in \Cref{app:exponential} that such results can be used to prove a balanced Voronoi partitioning bound with $\alpha = 1/2^{O(d)}$ for $d$-dimensional Euclidean space. Our \Cref{thm:main} improves this bound to $\Omega(1/d)$.

\medskip

\noindent\textbf{Efficient Opportunistically Navigable Graphs via Balanced Voronoi Partitions.}
Our results on balanced Voronoi partitions show that existing bounds for navigable graphs can be significantly improved, if one considers a slightly different search rule: \emph{opportunistic greedy search}. Here, the search  continues to move to the closest out-neighbor to a query $q \in X$, even if this neighbor is further from $q$ than the current vertex. The search terminates once $q$ is found (or if the current vertex has no out-neighbors). The search rule is in fact closer to common search rules used in practice, like beam search~\cite{AlJazzaziEtAl:2025}. We show:
%
%We consider the following variant of the search rule, applied to a graph whose vertices are labeled by points of $X$. A point may label more than one vertex.
%\begin{definition}[Opportunistic Greedy Search]
%Let $G=(V,E)$ be a directed graph with labels $\lambda:V\to X$. Given a target $q\in X$, \emph{opportunistic greedy search} terminates when it reaches a vertex $v$ with $\lambda(v)=q$. Otherwise, it moves to an out-neighbor $w$ minimizing $D(q,\lambda(w))$, with ties resolved by a rule specified as part of the construction. If no out-neighbor exists, the search terminates unsuccessfully.
%\end{definition}
%In particular, the search continues even when the next label is farther from $q$ than the current label. Our balanced partitioning theorem gives a simple construction for this rule.

\begin{theorem}[Efficient Opportunistically Navigable Graphs]
\label{thm:navigation-intro}
For every set $X$ of $n$ points in $\R^d$ under the Euclidean distance, there is a directed graph $G$ with vertices corresponding to points in $X$ and maximum out-degree two such that opportunistic greedy search reaches any target $q\in X$ from any starting vertex in $
 O(d\log n)$ 
steps and distance evaluations.
\end{theorem}
The construction of \Cref{thm:navigation-intro} is simple: it is just a search tree that recursively splits $X$ using balanced Voronoi partitions. The root has two children, corresponding to the points $a$ and $b$ for an $\alpha$-balanced Voronoi partition of $X$. When the query $q$ is closer to $a$ (i.e., $q \in S(a,b)$) the search proceeds to $a$. Otherwise, it proceeds to $b$. These nodes in turn have children that partition $S(a,b)$ and $S(b,a)$ respectively, and so on. The balance guarantee of \Cref{thm:main} means that the tree has depth at most $O(\log n/\alpha) = O(d \log n)$, bounding the routing time and number of distance evaluations. To allow starting from an arbitrary vertex, we simply add an edge from each leaf to the root. An unsuccessful first descent through the tree starting from an arbitrary point then returns to the root, after which the search descends the full tree and finds the query {$q$}. 

\Cref{thm:navigation-intro} significantly improves on existing results for navigable graphs with standard greedy search -- requiring just a logarithmic dependence on the dataset size $n$ and a linear dependence on the dimension. We remark that, via \Cref{thm:bounded-aspect-intro,thm:bounded-aspect-general-intro}, it directly extends to $\ell_p$ spaces, or other metric spaces with bounded SASs as well. An interesting open question is if such recursive partitioning techniques can give results for approximate nearest neighbor search when the query $q$ is not in the dataset. Haghiri et al.~\cite{HaghiriGhoshdastidarLuxburg:2017} study a very similar algorithm, but, as we will discuss in the next section, only {give} bounds under strong assumptions on the dataset.

%\smallskip

%\noindent \textbf{Connection to Comparison-Based Nearest Neighbor Search.}
%
%\Cam{NEed ot finish this section.}

\subsection{Related Work}

Below we discuss work in several different research areas related to our findings.

\smallskip

\noindent\textbf{Balanced Voronoi Partitions and Clusterings.} Not surprisingly, many works consider the construction of balanced Voronoi partitions or clusterings, where roughly equal numbers of points are assigned to each Voronoi cell or cluster center \cite{balzer2008capacity,pyatkin2017np,zhang2021improved}. Applications include political redistricting \cite{ricca2008weighted}, graphics \cite{balzer2009capacity}, and 
data analysis \cite{banerjee2006scalable,malinen2014balanced}. 
Typically, these works consider choosing $k \ge 2$ cluster centers, rather than exactly two as we do in \Cref{def:balancedPartition}. 
To the best of our knowledge, no existing work gives a balance guarantee for arbitrary point sets like  \Cref{thm:main}. Or, to give strong guarantees, prior work utilizes additional degrees of freedom, like placing weights on the cluster centers that impact their number of assigned points \cite{aurenhammer1998minkowski}.

Close in spirit to our work, \cite{honiden2009balancing,de2015interference,har2016space} ask how many centers must be chosen so that the cells in their Voronoi partition all contain an $O(1/k)$ fraction of the input points. Roughly speaking, these papers show that $\tilde O(k)$ centers suffice. However, the constants in their bounds depend exponentially on the dimension $d$. Further, perhaps surprisingly given our positive results for the $k =2$ case, this dependence is required: there exist point sets in $d$-dimensional Euclidean space such that $\Omega(\exp(d))$ centers must be chosen even for the maximum number of points in any Voronoi cell to be upper bounded by $|X|/4$ -- see \Cref{app:kcenter} for a proof. Given this  barrier, we focus on the $k = 2$ case in this paper, and on lower bounding the minimum fraction of points in a Voronoi cell by $1/\poly(d)$ -- i.e., on  upper bounding the maximum fraction by $(1-1/\poly(d))$.

\smallskip

\noindent \textbf{Centerpoint Theorem.} At a high level, our result on balanced Voronoi partitions is reminiscent of the \emph{centerpoint theorem}, which guarantees that for any set of points $X$ in $\R^d$, there is a point $x$ in the convex hull of $X$ such that \emph{every} hyperplane through $x$ divides $X$ into two roughly balanced sets -- with the smaller having size at least $|X|/(d+1)$. Centerpoints have applications in computational geometry, as they give a notion of a median in high dimensions \cite{clarkson1993approximating,jadhav1993computing}. A key difference is that the dividing hyperplane is only required to pass through the centerpoint, which may not itself be in $X$, rather than be the bisector between two points $a,b \in X$, as in a balanced Voronoi partition. Thus, we are not aware of any formal connection between the two concepts.

\smallskip

\noindent \textbf{Self-Approaching Curves.}
Our notion of self-approaching sequences can be thought of as the discrete analog of continuous \emph{self-approaching curves}, which have been studied in geometry \cite{rote1994curves,icking1999self}. However, this work uses a weaker notion than we do -- each consecutive point on the curve only needs to get closer to all subsequent points. Under this notion, there easily exist infinite point sets satisfying the property: e.g., any set of points along a straight line. Interestingly, work on self-approaching curves has been used in work on navigable graph spanners -- see e.g., \cite{alamdari2012self}. 

\smallskip

\noindent \textbf{Comparison-Based Nearest Neighbor Search.}
Our application to navigable graphs is closely related to work on \emph{comparison-based nearest neighbor search}, where the goal is to find a near neighbor to a query $q$, given only the ability to make distance-comparison queries (i.e., to check if $D(x_i,q) \le D(x_j,q)$ for dataset points $x_i,x_j$) \cite{GoyalLifshitsSchutze:2008,LifshitsZhang:2009,tschopp2011randomized,HaghiriGhoshdastidarLuxburg:2017,CohenAddadEtAl:2025}. Our algorithm falls under this framework, and is guaranteed to find $q$ exactly when $q \in X$. Prior works give bounds  even when $q \notin X$, by assuming upper bounds on the \emph{rank disorder} of the dataset \cite{GoyalLifshitsSchutze:2008,LifshitsZhang:2009,tschopp2011randomized}. Roughly speaking, they require that if $x$ is one of the closest neighbors to $y$ and $y$ is one of the closest neighbors to $z$, then $x$ is also a fairly close neighbor to $z$. Unfortunately, such a bound can fail even in one-dimensional Euclidean space. E.g., $x = 1$, $y = 2$, $z = 4$, and $n-3$ other points lying just above $6$. Then $x$ is the nearest neighbor of $y$ and $y$ is the nearest neighbor of $z$. But $x$ is the \emph{farthest} point from $z$ (at distance $3$ when all others have distance {$\approx 2$)}.

Very close to our navigable graph construction, Haghiri et al.~\cite{HaghiriGhoshdastidarLuxburg:2017} study a binary tree approach that recursively chooses two random points and assigns every other point to the closer one. They prove that their tree has height $O(\log n)$ with high probability, and give nearest neighbor search guarantees under a  \emph{strong expansion} condition. Again, this condition can fail even in one-dimensional Euclidean space. In particular, the condition implies that the number of points in $X$ lying within distance $2r$ of any $x \in X$ is bounded by a constant times the number lying within distance $r$. This is violated e.g., if we have one point at $x = 0$, $n-1$ points just above $1$, and take $r = 1$.

Finally, 
Cohen-Addad et al.~\cite{CohenAddadEtAl:2025} study comparison-based linear optimization, showing that a linear objective can be minimized over an $n$ point set using $O(k\log k\log n)$ comparisons, where $k$ is the set's \emph{conic dimension}, a variant of the \emph{inference dimension} of \cite{kane2017active}. Euclidean distance can be reframed as a linear objective by lifting $x\mapsto(x,\norm{x}_2^2)$ and rewriting $\norm{q-x}_2^2
 =\norm{q}_2^2+\langle (-2q,1),(x,\norm{x}_2^2) \rangle.$ The conic dimension is the maximum length of a sequence $(y_1,\ldots,y_k)$ of points from the set such that, for every $t\le k$, $y_t-y_1$ lies outside the cone generated by $\{y_j-y_i:1\le i<j<t\}$. While it is not bounded in general in low-dimensional Euclidean space, \cite{CohenAddadEtAl:2025} show that it is bounded by $O(d \log d)$ for points in $\{0,1\}^d$. They give bounds on a related approximate conic dimension notion for general  point sets with bounded aspect ratio.  
It would be interesting to understand if bounds on conic dimension or inference dimension are related to bounds on self-approaching sequence length -- the notions seem at least superficially related.

\section{Reduction from Balanced Voronoi Partitions to Bounded SASs}
\label{sec:reduction}

As discussed, the proofs of our balanced-partitioning results proceed in two steps. In this section, we  first prove the reduction from balanced partitioning to self-approaching sequences (\Cref{thm:reduction}): if a point set has no sufficiently
balanced Voronoi partition, then it contains a long
SAS. We then prove bounds on the length of SASs in $d$-dimensional Euclidean and $\ell_p$ spaces in \Cref{sec:euclidean-sas}. %Euclidean space has length $O(d^{5/2}\log d)$. Combining these
%two results proves \Cref{thm:main}.
%In this section, we prove the reduction of \Cref{thm:reduction}:

Throughout, we write $X = \{x_1,\ldots,x_n\}$ and break distance ties by ID, as in \Cref{def:balancedPartition}. Formally, for $x \in X$ and distinct $x_i,x_j \in X$, we say that $x$ \emph{prefers} $x_i$ to $x_j$ if $D(x,x_i) < D(x,x_j)$, or if $D(x,x_i) = D(x,x_j)$ and $i < j$. Note that, for each $x \in X$, the preferences of $x$ form a total order on $X$, in which $x$ itself comes first since $D(x,x) = 0$. For distinct $a,b \in X$, we write $S(a,b) = \{x \in X : x \text{ prefers } a \text{ to } b\}$. Note that 
$S(a,b)$ and $S(b,a)$ partition $X$, and $a \in S(a,b)$.

Our proof of \Cref{thm:reduction} uses the following simple proposition, which shows that to check if a sequence  is an SAS, it suffices to check the SAS condition for \emph{consecutive} pairs of points.
\begin{proposition}[SAS Consecutive Pair Characterization]\label{prop:consecutive}
Consider any metric space $(\mathcal X,D)$. A sequence of distinct points $(x_1,\ldots,x_\ell) \in \mathcal X$ is an SAS if and only if, for every $i \in \{1,\ldots,\ell-1\}$ and every $k \neq i$,
\begin{align}\label{eq:adjacent-pairs}
D(x_k,x_{i+1}) \le D(x_k,x_i).
\end{align}
\end{proposition}
\begin{proof}
If $(x_1,\ldots,x_\ell)$ is an SAS, then \eqref{eq:adjacent-pairs} is just the special case of \Cref{def:sas} with $j = i+1$. Conversely, suppose that \eqref{eq:adjacent-pairs} holds, and fix $i < j$ and $k \neq i$. If $k \notin \{i+1,\ldots,j-1\}$, then $k$ differs from the first index of each of the consecutive pairs $(x_{j-1},x_j), (x_{j-2},x_{j-1}), \ldots, (x_i,x_{i+1})$. So, applying \eqref{eq:adjacent-pairs} with base point $x_k$ to each of these pairs gives $D(x_k,x_j) \le D(x_k,x_{j-1}) \le \ldots \le D(x_k,x_i)$. If $i < k < j$, the same argument, applied to the consecutive pairs from $x_i$ to $x_{k-1}$ and from $x_{k+1}$ to $x_j$ gives $D(x_k,x_{k-1}) \le D(x_k,x_i)$ and  $D(x_k,x_j) \le D(x_k,x_{k+1})$. 

Further, applying \eqref{eq:adjacent-pairs} with base point $x_{k+1}$ to the pair $(x_{k-1},x_k)$, and  base point $x_{k-1}$ to the pair $(x_k,x_{k+1})$, gives
$
 D(x_k,x_{k+1}) \le D(x_{k+1},x_{k-1}) \le D(x_{k-1},x_k).$
Chaining these inequalities gives \[D(x_k,x_j) \le D(x_k,x_{k+1}) \le D(x_k,x_{k-1}) \le D(x_k,x_i),\] as required by \Cref{def:sas}.
\end{proof}

We now proceed with the proof of \Cref{thm:reduction}, which we restate below.

\begin{reptheorem}{thm:reduction}
Consider any metric space $(\mathcal X,D)$ and any set of points $X \subseteq \mathcal X$. Let $L_X$ denote the length of the longest SAS consisting of points in $X$. Then $X$ admits an $\alpha$-balanced Voronoi partition for $\alpha \ge 1/(4L_X)$.
\end{reptheorem}
\begin{proof}
%Let $D(\cdot,\cdot)$ denote the distance function associated with metric space $\mathcal X$.
Let {$n = |X|$}. % and recall that we assume $X$ is \emph{generic} -- so 
%$D(a,b)\ne D(a,c)$ for all $a,b,c\in X$ with $b \neq c$.
% \end{proof}
%
% \begin{proposition}[Bisectors from SASs]
% \label{prop:bisector-to-sas}
% \label{thm:voronoi-to-sas}
% Let $X$ be a generic $n$-point metric space.  Then
% \[
%  \max_{\{a,b\}\subseteq X}
%  \min\{|S(a,b)|,|S(b,a)|\}
%  \ge \ReductionConstant\,\frac{n}{\saslen(X)^2}.
% \]
% \end{proposition}
%
% \paragraph{Proof idea.}
% Orient each pair toward its smaller Voronoi side.  If every side is small,
% the resulting tournament is transitive.  A sparse auxiliary hypergraph then
% contains a large independent set, and the transitive order on that set is a
% self-approaching sequence.  The quantitative loss is quadratic in its
% length.
%
% \begin{proof}
Define  
\[
 M:=\max_{\{a,b\}\subseteq X}
       \min\{|S(a,b)|,|S(b,a)|\}.
\]
Our goal is to prove that $M \ge n/(4L_X)$. Note that if $M\ge n/3$,
the claim follows trivially since $L_X \ge 1$ and so $n/(4L_X) \le n/4$. So assume $M<n/3$.

We construct a tournament graph over points in $X$ as follows. For each unordered pair $\{a,b\}$, draw the arc $a\to b$ when
$|S(a,b)|\le M$ -- see \Cref{fig:tournament}(a).  Exactly one orientation is chosen: $S(a,b)$ and $S(b,a)$ partition $X$, so they cannot both have size $\le M<n/3$.

\begin{figure}[!b]
\vspace{1em}
\centering
\begin{tabular}{@{}c@{\hspace{1.5em}}c@{}}
\includegraphics[page=1]{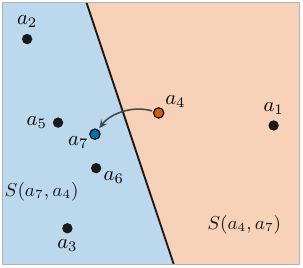} & \includegraphics[page=2]{graphics/fig-tournament.pdf} \\
\small (a) $|S(a_4,a_7)| = 2 < |S(a_7,a_4)| = 5$, so $a_4 \to a_7$ & \small (b) the tournament and the SAS $(a_1,a_2,a_6,a_7)$
\end{tabular}
\caption{The tournament graph constructed in the proof of \Cref{thm:reduction}. (a) Each edge $(a,b)$ is oriented towards the endpoint on the larger side of their Voronoi partition. (b) If there is no Voronoi partition with balance $\alpha \ge 1/3$, the tournament has no directed triangles and is transitive. Thus, the points can be placed in a total order that respects the tournament edges. The proof extracts a long SAS as a subsequence of this order. Here, the extracted SAS is $(a_1,a_2,a_6,a_7)$.}
\label{fig:tournament}
\end{figure}

Further, there can be no directed triangles in the tournament.  Indeed, if
$a\to b\to c\to a$, we must have $S(a,b) \cup S(b,c) \cup S(c,a) = X$; otherwise  $x \in X \setminus (S(a,b) \cup S(b,c) \cup S(c,a))$ would prefer $b$ to $a$, $c$ to $b$, and $a$ to $c$, which is impossible since the preferences of $x$ form a total order.
But we cannot have $S(a,b) \cup S(b,c) \cup S(c,a) = X$ since $|S(a,b) \cup S(b,c) \cup S(c,a)| \le 3M < n = |X|$, which gives a contradiction.

It is well known that a tournament
with no directed triangle is transitive, so we may relabel the points in $X$ as $a_1,a_2,\ldots, a_n$ so that
\[
 a_1\to a_2\to\cdots\to a_n.
\]
Since the tournament is transitive, whenever $i < j$, $a_i \to a_j$, and so $|S(a_i,a_j)|\le M$.

To prove the theorem, we will construct a subsequence of $(a_1,\ldots,a_n)$ that forms an SAS and has length at least $n/(4M)$ -- see \Cref{fig:tournament}(b). Rearranging will give that $M \ge n/(4L_X)$, as desired.

\smallskip

\noindent\textbf{A Simple Quadratic Bound.} We first note that sampling a set of roughly $\sqrt{n/M}$ random points already gives a weaker bound of  $M \ge n/(4L_X^2)$. Call a triple $(a_i,a_j,a_k)$ with $i<j$ and $k \neq i$ a \emph{violation} if $a_k \in S(a_i,a_j)$, i.e., if the SAS condition of \Cref{def:sas} is violated. By the ordering above, there are at most $M\binom{n}{2}$ violations. If we retain each point independently with probability $p = 1/\sqrt{Mn}$, then $pn = \sqrt{n/M}$ points are retained in expectation, while in expectation at most $p^3 M\binom{n}{2} \le \frac{1}{2}\sqrt{n/M}$ violations survive. Deleting one point from each surviving violation thus leaves, for some outcome, at least $\frac{1}{2}\sqrt{n/M}$ points among which there are no violations. In the order $a_1,\ldots,a_n$, these points form an SAS: for any such $a_i,a_j,a_k$ with $i<j$ and $k \neq i$, $a_k$ prefers $a_j$ to $a_i$, and so $D(a_k,a_j) \le D(a_k,a_i)$. So, $L_X \ge \frac{1}{2}\sqrt{n/M}$, which rearranges to $M \ge n/(4L_X^2)$.

To obtain a linear dependence on $L_X$, we instead only count violations between \emph{consecutive} retained points, which by \Cref{prop:consecutive} suffices to certify an SAS.

\smallskip

\noindent\textbf{Counting Consecutive Violations.} For a subsequence $R = (w_1,\ldots,w_m)$ of $(a_1,\ldots,a_n)$, define its number of consecutive violations as
\begin{align}\label{eq:violations}
 V(R) := \sum_{i=1}^{m-1} \left(|S(w_i,w_{i+1}) \cap R| - 1\right).
\end{align}
Each term counts the points of $R$, other than $w_i$ itself, that prefer $w_i$ to $w_{i+1}$. %Note that $w_i \in S(w_i,w_{i+1})$ and $w_{i+1} \notin S(w_i,w_{i+1})$, since every point prefers itself to any other point.
 If $V(R) = 0$, then for all $i \in \{1,\ldots,m-1\}$ and $k \neq i$, $w_k$ prefers $w_{i+1}$ to $w_i$, so $D(w_k,w_{i+1}) \le D(w_k,w_i)$. Thus $R$ is an SAS by \Cref{prop:consecutive}. 
 We next show that, in general, $R$ contains an SAS of length $|R| - V(R)$.

\smallskip

\noindent\textbf{Deleting Violations.} Suppose that $V(R) > 0$. Then there is some $i$ and some $w_k \in S(w_i,w_{i+1})$ with $k \neq i$. Since $w_{i+1} \notin S(w_i,w_{i+1})$, we also have $k \neq i+1$. Let $R'$ be $R$ with $w_k$ deleted. We claim that $V(R') \le V(R) - 1$. The pair $(w_i,w_{i+1})$ is still consecutive in $R'$, and its term in \eqref{eq:violations} decreases by at least one. The terms for all other consecutive pairs that do not involve $w_k$ cannot increase. Finally, if $1 < k < m$, the terms for the pairs $(w_{k-1},w_k)$ and $(w_k,w_{k+1})$ are replaced by a single term for the new consecutive pair $(w_{k-1},w_{k+1})$. We have
\[
 S(w_{k-1},w_{k+1}) \subseteq S(w_{k-1},w_k) \cup S(w_k,w_{k+1}),
\]
since otherwise some point would prefer $w_{k-1}$ to $w_{k+1}$, $w_k$ to $w_{k-1}$, and $w_{k+1}$ to $w_k$, which is impossible as each point's preferences form a total order. Since $w_k \in S(w_k,w_{k+1})$ and $w_k \in R$ but $w_k \notin R'$, it follows that
\[
 |S(w_{k-1},w_{k+1}) \cap R'| \le |S(w_{k-1},w_k) \cap R| + |S(w_k,w_{k+1}) \cap R| - 1.
\]
So, the new term is at most the sum of the two terms it replaces, and $V(R') \le V(R)-1$. Since $|R'| = |R| - 1$, we have $|R'| - V(R') \ge |R| - V(R)$. Repeating this deletion until no violations remain, we obtain an SAS of length $ \ge|R| - V(R)$. See \Cref{fig:violations} for an illustration of the process.

\begin{figure}[t]
\vspace{1em}
\centering
\begin{tabular}{@{}c@{\hspace{1.5em}}c@{}}
\includegraphics[page=1,scale=0.95]{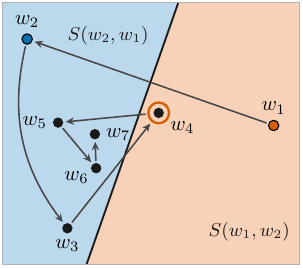} & \includegraphics[page=2,scale=0.95]{graphics/fig-violations.pdf} \\
\small (a) $w_4 \in S(w_1,w_2)$ & \small (b) after deleting $w_4$: $w_5 \in S(w_2,w_3)$ \\[1ex]
\includegraphics[page=3,scale=0.95]{graphics/fig-violations.pdf} & \includegraphics[page=4,scale=0.95]{graphics/fig-violations.pdf} \\
\small (c) after deleting $w_5$: $w_3 \in S(w_6,w_7)$ & \small (d) after deleting $w_3$: the SAS $(w_1,w_2,w_6,w_7)$
\end{tabular}
\caption{Deleting consecutive violations from a subsequence $R = (w_1,\ldots,w_m)$ of the total order $(a_1,\ldots,a_n)$ to construct an SAS. Here, $R$ consists of the full order of points in \Cref{fig:tournament}. (a)--(c) At each step, the Voronoi partition of one consecutive pair $(w_i,w_{i+1})$ has a  point $w_k$ with $k \neq i$ that lies in $S(w_i,w_{i+1})$, violating the SAS condition. This point is deleted, reducing the total number of violations by at least one, until no violations remain and the remaining points form an SAS.}
\label{fig:violations}
\end{figure}

\smallskip

\noindent\textbf{Finding a Good Subsequence.} We now show how to find a subsequence $R$ of $(a_1,\ldots,a_n)$ with $|R|-V(R) \ge n/(4M)$, giving us an SAS of the required length. 

Let $R$ be the random subsequence of $(a_1,\ldots,a_n)$ that retains each point independently with probability $q = \frac{1}{2M}$. We have $\mathbb E[|R|] = qn$. Consider any $i < j$. The points $a_i$ and $a_j$ are consecutive in $R$ exactly when both are sampled and none of $a_{i+1},\ldots,a_{j-1}$ are. If this happens, a point $a_k$ counted in the term of \eqref{eq:violations} for the consecutive pair $(a_i,a_j)$ must lie in $S(a_i,a_j)$, must be sampled, and must satisfy $k \notin \{i,i+1,\ldots,j-1,j\}$. Thus, since $|S(a_i,a_j)| \le M$ for all $i<j$,
\[
 \mathbb E[V(R)] \le \sum_{i<j} |S(a_i,a_j)| \cdot q^3 (1-q)^{j-i-1} \le Mq^3 \cdot n\sum_{g=0}^{\infty}(1-q)^{g} = Mq^2n.
\]
So, by linearity of expectation, $\mathbb E[|R| - V(R)] \ge qn - Mq^2 n = \frac{n}{2M} - \frac{n}{4M} = \frac{n}{4M}$. Thus, for some outcome, $|R| - V(R) \ge \frac{n}{4M}$, and so $X$ contains an SAS of length at least $\frac{n}{4M}$. That is, $L_X \ge \frac{n}{4M}$. Rearranging gives $M \ge \frac{n}{4L_X}$, as required.
\end{proof}

%\Cam{STOPPING HERE FOR NOW. }

% \begin{corollary}[Reduction principle]
% \label{cor:reduction-principle}
% If every SAS in a generic member $X$ of a metric family has length at most
% $L$, then some pair of sites has two strict Voronoi sides of size at least
% $9n/(64L^2)$.
% \end{corollary}

% \begin{proof}
% Apply \cref{thm:voronoi-to-sas} and use $\saslen(X)\le L$.
% \end{proof}

% !TEX root = ../main.tex

\section{Bounds on Self-Approaching Sequences}
\label{sec:euclidean-sas}

With the reduction of \Cref{thm:reduction} in place, we now turn to proving our SAS length upper bounds. In \Cref{sec:bounded-aspect} we prove \Cref{thm:bounded-aspect-general-intro}, which gives a general upper bound on the length of SASs with bounded aspect ratios in metric spaces that admit bounds on the size of nearly equilateral point sets. We apply this bound to give $\poly(d)$ upper bounds for Euclidean and $\ell_p$ SASs with bounded aspect ratio. In \Cref{sec:euclidean-poly} we show that for Euclidean space, the bounded aspect ratio assumption can be removed, yielding a weaker version of \Cref{thm:Ld} with a $\poly(d)$ upper bound for SASs in $d$-dimensional Euclidean space (\Cref{thm:Ld-poly}). Finally, in \Cref{sec:linear-sas} we show how to prove the tight $O(d)$ upper bound of \Cref{thm:Ld}, using an argument specialized to Euclidean space. 

%for Euclidean and $\ell_p$ spaces with bounded aspect ratio.  \Cref{thm:Ld}, which bounds the maximum length of an SAS in $d$-dimensional Euclidean space by $O(d)$. We first prove a weaker $\poly(d)$ bound (\Cref{thm:Ld-poly}) in \Cref{sec:bounded-aspect,sec:euclidean-poly}, since its proof is relatively simple, and introduces ideas that extend to $\ell_p$ spaces and general metric spaces with bounded aspect ratio. We then prove \Cref{thm:Ld} in \Cref{sec:linear-sas}.

%We now turn to the geometric part of the argument.  Two ingredients do most
%of the work: a packing bound for directions and a linear bound for sets whose
%pairwise distances lie in a sufficiently narrow interval.

%\Cam{STOPPED HERE FOR NOW.}

\subsection{Upper Bound for SASs with Bounded Aspect Ratio}\label{sec:bounded-aspect}

We start by proving \Cref{thm:bounded-aspect-general-intro}, which bounds the length of an SAS with bounded aspect ratio \eqref{eq:aspect} in any metric space that admits an upper bound on the size of a nearly equilateral point set. 
%In \Cref{sec:euclidean-poly}, we prove a weaker, polynomial version of \Cref{thm:Ld} by reducing the general case to this bounded aspect ratio case. %We restate \Cref{thm:bounded-aspect-general-intro} below:
\begin{samepage}
\begin{reptheorem}{thm:bounded-aspect-general-intro}
Consider any metric space $(\mathcal X,D)$ and suppose that, for some $\eps>0$, every set of points in this space with aspect ratio $\le 1+\eps$ has size at most $N_\eps$. For any $b>1$, every SAS $X$ in $(\mathcal X,D)$ with $\Delta(X) < b$ has length at most
\[
 N_\eps\cdot\left\lceil\log_{1+\eps}b\right\rceil.
\]
\end{reptheorem}
\end{samepage}
\begin{proof}
Let $(x_1,\ldots,x_\ell)$ be an SAS with aspect ratio less than $b$, and let 
$L_i=D(x_i,x_{i+1})$ for $1\le i<\ell$.  The SAS condition (\Cref{def:sas}), with
base point $x_i$ and ordered indices $i-1<i+1$ gives that 
$D(x_{i-1},x_i) \ge D(x_i,x_{i+1}).$
I.e., the $L_i$ are 
{non-increasing}.  

Let $m$ denote the minimum pairwise distance between points in the SAS, so by the aspect ratio bound, all pairwise distances lie in $[m,mb)$. In particular, $L_i \in [m,mb)$ for all $i$. Bucket the $L_i$ into $\lceil \log_{1+\eps} b \rceil$ buckets such that the $k^{th}$ bucket contains all $L_i$ lying in the range $[(1+\eps)^{k-1}m,(1+\eps)^k m)$. %each bucket contains $L_s,L_{s+1},\ldots, L_{t-1}$ with $L_$ 
Due to the monotonicity of the $L_i$, each bucket contains a contiguous sequence $L_s,L_{s+1},\ldots,L_{t-1}$ for some $1 \le s < t \le \ell$. We claim that, for all $i < j$ with $s\le i<j\le t$,
\begin{equation}
\label{eq:block-distance}
(1+\eps)^{k-1} m \le D(x_i,x_j)\le(1+\eps)^{k} m.
\end{equation}
I.e., the points $x_s,x_{s+1},\ldots,x_t$ have aspect ratio at most $1+\eps$. Assuming \eqref{eq:block-distance} holds, by our assumption on nearly equidistant points in our metric space, each bucket contains $\le N_\eps$ points. Since there are $\lceil \log_{1+\eps} b \rceil$ buckets, the total length of the SAS is bounded by $N_\eps \cdot \lceil \log_{1+\eps} b \rceil $, as required.

We now prove \eqref{eq:block-distance}.
For the upper bound, $j \ge i+1$, so, by the SAS condition with base point $x_i$, 
\begin{align*}
    D(x_i,x_j) \le D(x_i,x_{i+1}) = L_i \le (1+\eps)^k m.
\end{align*}
Similarly, for the lower bound, since $i \le j-1$, by the SAS condition with base point $x_j$,
\begin{align*}
    D(x_i,x_j) \ge D(x_{j-1},x_{j}) = L_{j-1} \ge (1+\eps)^{k-1} m.
\end{align*}
Together, the above inequalities give \eqref{eq:block-distance} and complete the proof.
\end{proof}

\subsubsection{Applications of \texorpdfstring{\Cref{thm:bounded-aspect-general-intro}}{Theorem \getrefnumber{thm:bounded-aspect-general-intro}} to Euclidean and \texorpdfstring{$\ell_p$}{ℓₚ} Spaces}

\Cref{thm:bounded-aspect-general-intro} reduces the problem of bounding the length of an SAS with bounded aspect ratio in some metric space to bounding the size of a (nearly) equilateral point set in that metric space. Such bounds have been studied in significant prior work, see e.g., \cite{alon2003equilateral,swanepoel2004equilateral,swanepoel2004problem,alon2009perturbed}. 
For example, it is well known that in Euclidean space, for $\eps = \Theta(1/\sqrt{d})$, we can bound $N_\eps = O(d)$. In particular, in \Cref{app:geometric} we use a result of \cite{alon2009perturbed} to show:
\begin{lemma}[Nearly Equilateral Point Sets -- Euclidean Space]
\label{lem:rough-equidistant-l2}
There are universal constants $c_1,c_2 >0$ such that if
$x_1,\ldots,x_N\in\R^d$ satisfy
\[
 1\le\norm{x_i-x_j}_2\le1+\frac{c_1}{\sqrt d}\quad \text{ for all }i \neq j
\]
then $N \le c_2 d$.
\end{lemma}
Plugging \Cref{lem:rough-equidistant-l2} into {\Cref{thm:bounded-aspect-general-intro}} gives:
\begin{corollary}[Bounded Aspect Ratio SAS -- Euclidean Space]
\label{cor:b-bounded-euclid}
For  any $b> 1$, every SAS in $\R^d$ under the Euclidean distance with aspect ratio  smaller than $b$
has length $O(d^{3/2} \lceil \log b \rceil )$.
\end{corollary}

In \Cref{app:geometric} we also prove a bound on nearly equilateral sets in $\ell_p^d$ for any $p \ge 1$. This bound follows an approach of Alon and Pudl\'ak \cite{alon2003equilateral}, which focuses on the exactly equilateral setting.
\begin{lemma}[Nearly Equilateral Point Sets -- $\ell_p$ Spaces]
\label{lem:rough-equidistant-lp}
For any fixed $1\le p<\infty$, let $q_p = \frac{2p+2}{2p-1} \le 4$. There exist constants $c_{p,1}$ and $c_{p,2}$ (depending on $p$) such that, if
$x_1,\ldots,x_N\in \R^d$ satisfy
\[
 1\le\norm{x_i-x_j}_p\le1+c_{p,1} \cdot d^{-q_p/2} \quad \text{ for all }i \neq j,
\]
then $N \le c_{p,2} \cdot d^{q_p}$.
\end{lemma}
Plugging \Cref{lem:rough-equidistant-lp} into {\Cref{thm:bounded-aspect-general-intro}} gives:
\begin{corollary}[Bounded Aspect Ratio SAS -- $\ell_p$ Space]
\label{cor:b-bounded-lp}
For  any $b> 1$, every SAS in $\R^d$ under the $\ell_p$ norm for fixed $1 \le p < \infty$ with aspect ratio  smaller than $b$
has length $O_p(d^{3q_p/2} \lceil \log b \rceil )$, where $q_p = \frac{2p+2}{2p-1} \le 4$.
\end{corollary}

Via \Cref{thm:reduction}, \Cref{cor:b-bounded-euclid,cor:b-bounded-lp} imply corresponding $1/\poly(d)$ balanced Voronoi partition bounds for point sets in Euclidean or $\ell_p$ spaces with bounded aspect ratio (since any SAS consisting of points in a set $X$ has aspect ratio at most $\Delta(X)$). \Cref{cor:b-bounded-lp} in particular gives the SAS length bound and  balanced partitioning bound stated in \Cref{thm:bounded-aspect-intro}.
% \paragraph{Proof idea.}
% Consecutive edge lengths decrease along an SAS.  Group comparable edges
% together; the sequence condition then forces all pairwise distances in the
% corresponding vertex block to be comparable as well.  The roughly
% equidistant-set lemma bounds each block.

\subsection{A Polynomial Bound for General SASs in Euclidean Space}\label{sec:euclidean-poly}

We now prove a weaker version of \Cref{thm:Ld} giving a $\poly(d)$ upper bound by reducing the general case in Euclidean space to the bounded aspect ratio case analyzed in {\Cref{thm:bounded-aspect-general-intro}} and \Cref{cor:b-bounded-euclid} of the previous section. Our proof will rely on just one basic fact about $d$-dimensional Euclidean space, proven for completeness in \Cref{app:geometric}: there can be at most $O(d)$ unit vectors with pairwise inner products bounded by $1/(2d)$ -- i.e., that have pairwise angles close to $90^\circ$ or greater.

\begin{lemma}[Direction packing]\label{lem:direction-packing}
If unit vectors $u_1,\ldots,u_N\in\R^d$ satisfy
$\langle u_i,u_j \rangle\le 1/(2d)$ for all $i\ne j$, then $N\le4d$.
\end{lemma}

With this lemma in place, we can show:

\begin{theorem}[Polynomial Euclidean SAS Bound]\label{thm:Ld-poly}
Let $L_d$ denote the maximum possible length of an SAS in $\R^d$ under the Euclidean distance. There is a universal constant \(C\) such that, for every \(d\),
\[
        L_d \le C \cdot d^{2.5} \cdot \log(d+1).
\]
\end{theorem}
\begin{proof}
We assume $d > 1$ in the proof. For $d = 1$, the bound follows from the bound for $d = 2$.
Let $(x_1,\ldots,x_\ell)$ be an SAS. Translate the points so that $x_\ell=0$ and then scale them with a common scaling so that $\norm{x_{\ell-1}}_2 = 1.$ In this way, by the SAS condition applied with base point $x_\ell$, we have $\norm{x_1}_2 \ge \norm{x_2}_2 \ge  \ldots \ge \norm{x_{\ell-1}}_2 = 1.$

Let $\bar x_i = x_i/\norm{x_i}_2$. For any $i<j<\ell$, applying the SAS condition with base point $x_i$
and the ordered pair $j<\ell$ gives
\begin{align}\label{eq:anchor1}
\norm{x_i}_2^2 \le \norm{x_i-x_j}_2^2 = \norm{x_i}_2^2+\norm{x_j}_2^2-2 \cdot \norm{x_i}_2 \cdot \norm{x_j}_2 \cdot \langle \bar x_i,\bar x_j \rangle.
\end{align}
Rearranging gives
\begin{equation}
\label{eq:anchor}
 \langle \bar x_i,\bar x_j \rangle \le \frac{\norm{x_j}_2}{2\norm{x_i}_2}.
\end{equation}
Now, bucket the points so that the $k^{th}$ bucket contains all $x_i$ with {$\norm{x_i}_2 \in [d^{k-1},d^k)$}. The points in one bucket form an SAS with aspect ratio $< 2d$. Indeed as in \eqref{eq:anchor1}, for $i<j$, we have $\norm{x_i-x_j}_2\ge\norm{x_i}_2 \ge d^{k-1}$. On the other hand, the triangle inequality gives
$\norm{x_i-x_j}_2 \le \norm{x_i}_2+ \norm{x_j}_2 <2d^{k}$.  Thus, applying \Cref{cor:b-bounded-euclid}, each bucket has
$O(d^{3/2}\log d)$ points.

It remains to bound the number of nonempty buckets.  Select one representative point $x_i$ from
each nonempty even bucket.  Any two selected points $x_i,x_j$ have norms differing by a factor of {greater than}
$d$. So \eqref{eq:anchor} gives
$\langle \bar x_i,\bar x_j \rangle<1/(2d)$.  Thus, by \Cref{lem:direction-packing}, there are at
most $4d$ nonempty even buckets.  The same argument applies to the odd
buckets.  Multiplying the number of buckets by the bound within each bucket
proves {$L_d=O(d^{2.5}\log d)$}.
\end{proof}

\subsection{A Linear Bound for SASs in Euclidean Space}\label{sec:linear-sas}

We now prove \Cref{thm:Ld}, which improves the $\poly(d)$ Euclidean SAS bound of \Cref{thm:Ld-poly} to linear in $d$. As discussed, {$L_d \ge 2d$}, so this bound is tight up to a constant factor.
\begin{reptheorem}{thm:Ld}
Let $L_d$ denote the maximum possible length of an SAS in $\R^d$ under the Euclidean distance. For every $d$, $L_d \le {8d}.$
\end{reptheorem}

Our proof considers the successive differences $x_{i+1}-x_i$ between points in the SAS, arguing that they are nearly orthogonal on average. This captures the intuition that the points in an SAS cannot keep moving in the same direction while getting closer to all previous points. Formally:

\begin{lemma}[Successive SAS Differences are Nearly Orthogonal]\label{lem:step-directions}
Let $(x_1,\ldots,x_\ell)$ be an SAS in $\R^d$. For every $i \in \{1,\ldots, \ell-1\}$, let $u_i:=(x_{i+1}-x_i)/\norm{x_{i+1}-x_i}_2$ denote the unit normalized difference between the $(i+1)^{th}$ and $i^{th}$ points in the sequence. Then for every $i \in \{1,\ldots,\ell-1 \}$,
\[
 \sum_{j=i+1}^{\ell-1}\langle u_i,u_j \rangle^2\le {3}.
\]
\end{lemma}

\Cref{lem:step-directions} nearly directly yields \Cref{thm:Ld} via a linear algebraic argument, very similar to our bound on the number of unit vectors in $\R^d$ with pairwise angles close to $90^\circ$ or greater in \Cref{lem:direction-packing}. In \Cref{lem:direction-packing} we have an upper bound on every $\langle u_i,u_j \rangle$. In \Cref{lem:step-directions} we have a weaker aggregate bound, but the same general idea applies. 
\begin{proof}[Proof of \Cref{thm:Ld}]
Let $(x_1,\ldots,x_\ell)$ be an SAS in $\R^d$ and assume $\ell \ge 3$. Otherwise, $\ell \le {8d}$ trivially. Let $G \in \R^{(\ell-1) \times (\ell-1)}$ be the Gram matrix with $G_{ij} = \langle u_i,u_j \rangle$, where $u_i,u_j$ are the normalized differences defined in \Cref{lem:step-directions}. As in the proof of \Cref{lem:direction-packing} (see \eqref{eq:g3}), since $G$ is positive semidefinite with $\rank(G) \le d$ and $\tr(G) = \ell-1$, we have $\norm{G}_F^2 \ge (\ell-1)^2/d$. On the other hand, by \Cref{lem:step-directions},
\[
 \norm{G}_F^2
 = (\ell-1) + 2\sum_{i=1}^{\ell-1}\sum_{j=i+1}^{\ell-1}\langle u_i,u_j \rangle^2
 \le (\ell-1) + 2 \cdot {3} (\ell-1)
 = {7(\ell-1)}.
\]
{Combining the upper and lower bounds on $\norm{G}_F^2$ gives $\ell \le 7d+1 \le 8d$.}
\end{proof}

\subsubsection{Proof of \texorpdfstring{\Cref{lem:step-directions}}{Lemma \getrefnumber{lem:step-directions}}}

It remains to prove \Cref{lem:step-directions}. By the SAS condition, every step $x_j \to x_{j+1}$ moves closer to each earlier point $x_i$ with $i < j$. In other words, if we consider the bisector of $x_j$ and $x_{j+1}$ (the hyperplane of points equidistant from $x_j$ and $x_{j+1}$), with normal vector $u_j$, each earlier point lies on the $x_{j+1}$ side of this hyperplane.

\begin{figure}[t]
\centering
\includegraphics{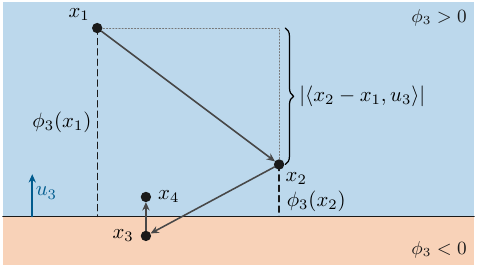}
\caption{An SAS $(x_1,x_2,x_3,x_4)$ in $\R^2$, the bisector of $x_3$ and $x_4$, and the unit normal $u_3$. By the SAS condition, the signed distances $\phi_3(x_1)$ and $\phi_3(x_2)$ to the bisector (\Cref{def:bisector-dist}) are nonnegative. They differ by $|\langle x_2-x_1,u_3\rangle| = \norm{x_2-x_1}_2 \cdot |\langle u_1,u_3\rangle|$, so if $|\langle u_1,u_3\rangle|$ is large, the bisector must be far from at least one of $x_1$ and $x_2$.}
\label{fig:bisector-dist}
\end{figure}

If we consider an earlier $i < j-1$, the (signed) distances of $x_i$ and $x_{i+1}$ to the $j^{th}$ bisector differ by exactly $\langle x_{i+1} - x_i,u_j \rangle = \norm{x_{i+1} - x_i}_2 \cdot \langle u_i,u_j \rangle$. Since both are nonnegative by the SAS condition, one of them is at least $\norm{x_{i+1} - x_i}_2 \cdot |\langle u_i,u_j \rangle|$. That is, if $|\langle u_i,u_j \rangle|$ is large, the $j^{th}$ bisector must be far from either $x_i$ or $x_{i+1}$.  See \Cref{fig:bisector-dist}. However, the bisectors of a sequence that keeps getting closer to a point $a$ cannot continually stay far from $a$: in \Cref{lem:bisector-sum} below we show that the squared distances from $a$ to successive bisectors sum to at most the squared distance to the first point. This implies that only $O(1)$ later steps can have large inner product with $u_i$, yielding  \Cref{lem:step-directions}.

Our proof repeatedly uses  the signed distance from a point to each bisector of a sequence, which we define notation for below.
\begin{definition}[Signed Bisector Distance]\label{def:bisector-dist}
Let $(x_1,\ldots,x_\ell)$ be a sequence of distinct points in $\R^d$. For each $i \in \{1,\ldots,\ell-1\}$, let $L_i := \norm{x_{i+1}-x_i}_2$ denote the distance between the $i^{th}$ and $(i+1)^{th}$ points, so that $u_i = (x_{i+1}-x_i)/L_i$, and define $\phi_i:\R^d \to \R$ by
\[
 \phi_i(a):=\frac{\norm{a-x_i}_2^2-\norm{a-x_{i+1}}_2^2}{2L_i}.
\]
\end{definition}
Expanding the squared norms gives
\begin{equation}\label{eq:phi}
 \phi_i(a)=\langle a-x_i,u_i \rangle-\frac{L_i}{2} = \left\langle a-\frac{x_i+x_{i+1}}{2},u_i \right\rangle.
\end{equation}
Since the bisector of $x_i$ and $x_{i+1}$ is the hyperplane through $(x_i+x_{i+1})/2$ with unit normal $u_i$, $\phi_i(a)$ is exactly the signed distance from $a$ to this bisector, which is positive on the side of $x_{i+1}$. %In particular, $\phi_i$ is affine with gradient $u_i$, and $\phi_i(a) \ge 0$ exactly when $a$ is at least as close to $x_{i+1}$ as to $x_i$. 
%So, a point $a$ gets (weakly) closer to each successive point of the sequence, i.e., $\norm{a-x_1}_2 \ge \cdots \ge \norm{a-x_\ell}_2$, if and only if $\phi_i(a) \ge 0$ for all $i \in \{1,\ldots,\ell-1\}$.

%Our main tool is the following bound on the distances from such a point $a$ to the bisectors of the sequence, which we prove via a potential function argument.

With this notation in place, we now show that if an SAS gets continually closer to a point $a$, then the sum of squared distances from $a$ to the sequence's bisectors is bounded.

\begin{lemma}[Bisector Distance Bound]\label{lem:bisector-sum}
Let $(x_1,\ldots,x_\ell)$ be an SAS in $\R^d$, and let $a\in\R^d$ satisfy $\norm{a-x_1}_2\ge\norm{a-x_2}_2\ge\cdots\ge\norm{a-x_\ell}_2$. Then
\[
 \sum_{i=1}^{\ell-1}\phi_i(a)^2\le\norm{a-x_1}_2^2.
\]
\end{lemma}
\begin{proof}
Following \Cref{def:bisector-dist}, we have $\phi_i(a) \ge 0$ exactly when $\norm{a-x_{i+1}}_2 \le \norm{a-x_i}_2$. So by the assumption that the sequence approaches $a$, we have that $\phi_i(a) \ge 0$ for all $i \in \{1,\ldots,\ell-1\}$.

We proceed by induction on $\ell$. If $\ell = 1$, the sum is empty. If $\ell=2$, then Cauchy-Schwarz and \eqref{eq:phi} give $\phi_1(a) = \langle a-x_1,u_1 \rangle - L_1/2 \le \norm{a-x_1}_2$, as required. Now, let $\ell\ge3$, and assume that the lemma holds for all SASs of length $\ell-1$.

\medskip
\noindent\textbf{Case 1: $\phi_k(a)=0$ for some $k \in \{1,\ldots,\ell-1\}$.} Removing $x_k$ from the sequence gives an SAS of length $\ell-1$ (any subsequence of an SAS is an SAS), and $a$ still gets weakly closer to each successive point of this sequence. If $k=1$, then $\norm{a-x_1}_2 = \norm{a-x_2}_2$. So, since $\phi_1(a) = 0$, applying the inductive hypothesis to $(x_2,\ldots,x_\ell)$ gives
\[
 \sum_{i=1}^{\ell-1}\phi_i(a)^2 = \sum_{i=2}^{\ell-1}\phi_i(a)^2\le\norm{a-x_2}_2^2 = \norm{a-x_1}_2^2.
\]
If $k \ge 2$, the shortened sequence $(x_1,\ldots,x_{k-1},x_{k+1},\ldots,x_\ell)$ has the same first point $x_1$. Its bisector distances at $a$ are $\phi_i(a)$ for $i \notin \{k-1,k\}$, together with
\[
 \phi'(a) := \frac{\norm{a-x_{k-1}}_2^2-\norm{a-x_{k+1}}_2^2}{2\norm{x_{k+1}-x_{k-1}}_2}
\]
for the new consecutive pair $(x_{k-1},x_{k+1})$. Since $\phi_k(a) = 0$, $\norm{a-x_{k+1}}_2 = \norm{a-x_k}_2$, so $\phi'(a)$ has the same numerator as $\phi_{k-1}(a)$. Further, $\phi'(a)$ has a smaller (or equal) denominator, since the SAS condition with base point $x_{k-1}$ gives $\norm{x_{k+1}-x_{k-1}}_2 \le \norm{x_k-x_{k-1}}_2 = L_{k-1}$. Thus, $\phi_{k-1}(a) \le \phi'(a)$. So, using that $\phi_k(a) = 0$ and applying the inductive hypothesis to the shortened sequence,
\[
 \sum_{i=1}^{\ell-1}\phi_i(a)^2 = \sum_{i \notin \{k-1,k\}}\phi_i(a)^2 + \phi_{k-1}(a)^2 \le \sum_{i \notin \{k-1,k\}}\phi_i(a)^2 + \phi'(a)^2 \le \norm{a-x_1}_2^2.
\]

\medskip
\noindent\textbf{Case 2: $\phi_i(a)>0$ for all $i \in \{1,\ldots,\ell-1\}$.} In this case, we move $a$ along the direction $u_1$ until we reach a point covered by Case 1. In particular, for $t \in \R$, let $a_t := a+tu_1$. By \eqref{eq:phi}, $\phi_i(a_t) = \phi_i(a)+t\langle u_1,u_i \rangle$. Let $T$ be the set of $t$ for which $a_t$ satisfies the assumption of the lemma, i.e., the sequence approaches $a_t$ and so $\phi_i(a_t) \ge 0$ for all $i$. Each of these conditions defines a closed half-line (or all of $\R$), so $T$ is a closed interval containing $0$. Further, $T$ is bounded. Since $\phi_1(a_t)=\phi_1(a)+t$, every $t \in T$ satisfies $t\ge-\phi_1(a)$. Since $\phi_2(a_t)=\phi_2(a)+t\langle u_1,u_2 \rangle$, and since the SAS condition with base point $x_1$ gives
\[
 L_1^2 = \norm{x_2-x_1}_2^2 \ge \norm{x_3-x_1}_2^2 = \norm{L_1u_1+L_2u_2}_2^2 = L_1^2+L_2^2+2L_1L_2\langle u_1,u_2 \rangle,
\]
we have $\langle u_1,u_2 \rangle\le-L_2/(2L_1)<0$, and so every $t \in T$ satisfies $t\le\phi_2(a)/|\langle u_1,u_2 \rangle|$. Let $t_- \le 0 \le t_+$ be the endpoints of $T$. At each endpoint, all $\phi_i(a_t)$ are nonnegative and, by continuity, at least one of them is equal to $0$. So by Case 1, the lemma holds for $a_{t_-}$ and $a_{t_+}$.

To extend this to $a = a_0$, consider the gap between the two sides of the desired inequality as we vary $t$:
\[
 g(t) := \sum_{i=1}^{\ell-1}\phi_i(a_t)^2-\norm{a_t-x_1}_2^2.
\]
We have shown that $g(t_-) \le 0$ and $g(t_+) \le 0$, and must show that $g(0) \le 0$. This follows since $g$ is convex. In particular, $\frac{\mathrm{d} \phi_i(a_t)}{\mathrm{d} t} = \langle u_1,u_i \rangle$. So %and $\langle u_1,u_1 \rangle = \norm{u_1}_2^2 = 1$. So,
\[
 g''(t) = 2\sum_{i=1}^{\ell-1}\langle u_1,u_i \rangle^2 - 2\norm{u_1}_2^2 = 2\sum_{i=2}^{\ell-1}\langle u_1,u_i \rangle^2\ge0,
\]
and thus $g(0) \le \max\{g(t_-),g(t_+)\} \le 0$.
\end{proof}

We can now prove \Cref{lem:step-directions} using \Cref{lem:bisector-sum}.
\begin{proof}[Proof of \Cref{lem:step-directions}]
For $i = \ell-1$, the sum is empty. So fix $i \le \ell-2$. Since the lemma statement is invariant to rescaling the points, we may assume that $L_i = 1$, so that $x_{i+1}-x_i = u_i$. Note that by the SAS condition with base point $x_{i+1}$, $L_{i+1} = \norm{x_{i+2}-x_{i+1}}_2 \le \norm{x_i-x_{i+1}}_2 = 1$.

%\medskip
%\noindent\textbf{The $j=i+1$ term.} Since $u_i$ and $u_{i+1}$ are unit vectors, $\langle u_i,u_{i+1} \rangle^2 \le 1$.
%\medskip
%\noindent\textbf{The $j\ge i+2$ terms.} 

By \eqref{eq:phi} we can write:
\[
 \langle u_i,u_j \rangle=\langle x_{i+1}-x_i,u_j \rangle=\phi_j(x_{i+1})-\phi_j(x_i).
\]
By the SAS condition, $(x_{i+2},\ldots,x_\ell)$ approach both $x_i$ and $x_{i+1}$. So $\phi_j(x_i),\phi_j(x_{i+1})\ge0$ for all $j\ge i+2$, and since $(\alpha-\beta)^2\le\alpha^2+\beta^2$ for $\alpha,\beta\ge0$,
\begin{align}\label{eq:squareSum}
 \langle u_i,u_j \rangle^2\le\phi_j(x_i)^2+\phi_j(x_{i+1})^2.
\end{align}
We bound the sums of these terms using \Cref{lem:bisector-sum}. We first apply the lemma with $a = x_i$ and SAS $(x_{i+1},\ldots, x_\ell)$, giving
\[
 \sum_{j=i+2}^{\ell-1}\phi_j(x_i)^2 \le \sum_{j=i+1}^{\ell-1}\phi_j(x_i)^2 \le \norm{x_i-x_{i+1}}_2^2 = 1.
\]
Similarly, applying \Cref{lem:bisector-sum} with $a = x_{i+1}$ and SAS $(x_{i+2},\ldots,x_\ell)$ gives
\[
 \sum_{j=i+2}^{\ell-1}\phi_j(x_{i+1})^2\le\norm{x_{i+1}-x_{i+2}}_2^2 = L_{i+1}^2 \le 1.
\]

\medskip
\noindent Plugging these bounds back into \eqref{eq:squareSum}, and noting that trivially $\langle u_i,u_{i+1} \rangle^2 \le 1$ by Cauchy-Schwarz,
\[
 \sum_{j=i+1}^{\ell-1}\langle u_i,u_j \rangle^2 \le 1 + 1 + 1 = 3. \qedhere
\]
\end{proof}

\section{Efficient Navigable Graphs Under Opportunistic Greedy Search}
\label{sec:navigation}

We now prove  \Cref{thm:navigation-intro}, our result on efficiently navigable graphs via balanced Voronoi partitions. %Throughout, $X$ is a set of $n \ge 2$ points in a metric space $(\mathcal X,D)$, and $G$ is a directed graph whose vertices correspond to the points in $X$. We identify each vertex with its point. 
As discussed in \Cref{sec:navigationIntro}, we will construct graphs that are navigable under an `opportunistic' variant of greedy search, that always moves to the closest out-neighbor of the current vertex, even if this neighbor is further from the query than the current vertex.

\begin{definition}[Opportunistic Greedy Search]\label{def:opportunistic}
Let $X = \{x_1,\ldots,x_n\}$ be a set of $n$ points in a metric space $(\mathcal X,D)$, and let $G$ be a directed graph with vertices corresponding to the points in $X$. Given a query $q \in X$ and a starting vertex $s \in X$, \emph{opportunistic greedy search} on $G$ starts at $v = s$ and repeats the following. If $v = q$, it terminates successfully. If $v \neq q$ has no out-neighbors in $G$, it terminates unsuccessfully. Otherwise, it moves to the out-neighbor $w$ of $v$ minimizing $D(q,w)$, with ties broken by id.
%We say that $G$ is \emph{opportunistically navigable} if, for every query $q \in X$ and every starting vertex $s \in X$, opportunistic greedy search on $G$ terminates successfully, i.e., reaches $q$.
\end{definition}
Note that, as defined, the search may never terminate. It can be made to always terminate if it only considers out-neighbors that were previously unvisited, or simply sets a maximum step bound. %Our graphs will be navigable under this variant as well. 
%In the terminology of \Cref{sec:reduction}, the search moves to the out-neighbor that $q$ prefers. Note that the search does not need to track which vertices it has visited. Each step evaluates the distance from $q$ to each out-neighbor of the current vertex.

\subsection{Balanced Voronoi Partition Trees}

As described in \Cref{sec:navigationIntro}, our search graph $G$ will be based on a tree that recursively splits $X$ using balanced Voronoi partitions. %As in \Cref{def:balancedPartition,def:opportunistic}, distance ties are broken by id, so each balanced partition is a true partition. 
Recall from \Cref{sec:reduction} that $x$ \emph{prefers} $a$ to $b$ if $D(x,a) < D(x,b)$, or if $D(x,a) = D(x,b)$ and $a$ has a smaller id than $b$, and that $S(a,b)$ is the set of points in $X$ that prefer $a$ to $b$.
For $Y \subseteq X$ and distinct $a,b \in Y$, let $S^Y(a,b) := S(a,b) \cap Y$ be the set of points in $Y$ that prefer $a$ to $b$. $S^Y(a,b)$ and $S^Y(b,a)$ partition $Y$. %Note that $S^Y(a,b)$ is exactly the set $S(a,b)$ defined with respect to the point set $Y$, with ids inherited from $X$. %, and $a \in S^Y_\prec(a,b)$. 
We construct our routing tree as follows.
\begin{definition}[Voronoi Routing Tree]\label{def:tree}
Let $X = \{x_1,\ldots,x_n\}$ be a set of $n$ points in a metric space $(\mathcal X,D)$. A \emph{Voronoi routing tree} $T$ is a rooted tree with vertex set $X$, where each vertex $v$ is associated with a set $Y_v \subseteq X \setminus \{v\}$ of its descendants. 
Pick an arbitrary root $r \in X$ and set $Y_r = X \setminus \{r\}$. Then, recursively, for each vertex $v$:
\begin{itemize}
 \item If $Y_v = \emptyset$, $v$ is a leaf.
 \item If $Y_v = \{x\}$, $v$ has a single child $x$, with $Y_x = \emptyset$ (i.e., $x$ is a leaf).
 \item If $|Y_v| \ge 2$, choose distinct $a,b \in Y_v$ maximizing $\min\{|S^{Y_v}(a,b)|,|S^{Y_v}(b,a)|\}$. Then $v$ has children $a$ and $b$, with $Y_a = S^{Y_v}(a,b) \setminus \{a\}$ and $Y_b = S^{Y_v}(b,a) \setminus \{b\}$.
\end{itemize}
 The corresponding Voronoi search graph $G$ is identical to ${T}$ but with an edge from each leaf to the root $r$.
\end{definition}
%By induction, the subtree rooted at each $v$ contains exactly $\{v\} \cup Y_v$, so every point of $X$ appears exactly once in $\mathcal T$. Every vertex of $G$ has out-degree at most two.

%\subsection{Routing Guarantee}

It is not hard to argue that opportunistic greedy search on the Voronoi search graph $G$ will reach any target $q \in X$ starting from any start node, within a number of steps bounded by the depth of $T$. In particular: 
\begin{lemma}[Voronoi Search Graph Routing Guarantee]\label{lem:routing}
Let $T$ and $G$ be as in \Cref{def:tree}, and let $h$ be the depth of $T$. For any query $q \in X$ and starting vertex $s \in X$, opportunistic greedy search (\Cref{def:opportunistic}) on $G$ reaches $q$ within $2h+1$ steps, each requiring at most two distance evaluations.
\end{lemma}
\begin{proof}
We first show that if the search is at a vertex $v$ with $q \in \{v\} \cup Y_v$, then it follows the path from $v$ to $q$ in $T$. If $v = q$, we are done. Otherwise $q \in Y_v$, so $v$ is not a leaf and its out-neighbors are its children. If $v$ has a single child, that child is $q$, and the search moves to it. Otherwise, since $S^{Y_v}(a,b)$ and $S^{Y_v}(b,a)$ partition $Y_v$, assume without loss of generality that $q \in S^{Y_v}(a,b)$. Then $q$ prefers $a$ to $b$. So, the search moves to $a$, the next vertex on the path to $q$. Since $q \in S^{Y_v}(a,b) = \{a\} \cup Y_a$, the same holds at $a$, and by induction, the search follows the path to $q$.

By the above, if $q \in \{s\} \cup Y_s$, the search reaches $q$ within $h$ steps. Otherwise, the search first descends $T$ from $s$, since the out-neighbors of any non-leaf vertex are its children. As $q$ is not in the subtree rooted at $s$, within $h$ steps the search reaches a leaf $\ell \neq q$, and then moves to the only out-neighbor of $\ell$, the root $r$. Since $q \in \{r\} \cup Y_r = X$, the search then follows the path from $r$ to $q$, reaching $q$ within $h$ further steps -- for at most $2h+1$ steps in total. Finally, since every vertex of $G$ has out-degree at most two, each step requires at most two distance evaluations.
\end{proof}

To apply \Cref{lem:routing} and give our final bound on efficient greedy routing (\Cref{thm:navigation-intro}), it remains to bound the depth of $T$.
\begin{lemma}[Balanced Voronoi Tree Depth Bound]\label{lem:tree-depth}
Let $T$ be as in \Cref{def:tree} and 
let $\alpha > 0$ be such that every $Y \subseteq X$ with $|Y| \ge 2$ contains distinct $a,b$ with
\begin{align}\label{eq:ordered-balance}
 \min\{|S^Y(a,b)|,|S^Y(b,a)|\} \ge \alpha|Y|.
\end{align}
Then $T$ has depth at most $\lceil \ln n/\alpha \rceil$.
\end{lemma}
\begin{proof}
Let $x$ be a child of $v$. Either $Y_x = \emptyset$, or $x \in \{a,b\}$. Assume $x = a$ without loss of generality. In this case, since by \Cref{def:tree}, $a$ and $b$ are chosen to maximize the left-hand side of \eqref{eq:ordered-balance}, we have $|S^{Y_v}(b,a)| \ge \alpha|Y_v|$, and so $|Y_x| = |S^{Y_v}(a,b)| -1 \le (1-\alpha)|Y_v| -1 < (1-\alpha) |Y_v|$. So, every vertex $v$ at depth $k$ has $|Y_v| \le (1-\alpha)^k (n-1) < e^{-\alpha k} n$. For $k \ge \lceil \ln n/\alpha \rceil$, this is at most $1$, so $Y_v = \emptyset$ and $v$ is a leaf.
\end{proof}

Combining \Cref{lem:routing,lem:tree-depth} with \Cref{thm:reduction,thm:Ld}, we can now prove \Cref{thm:navigation-intro}:
\begin{reptheorem}{thm:navigation-intro}
For every set $X$ of $n$ points in $\R^d$ under the Euclidean distance, there is a directed graph $G$ with vertices corresponding to points in $X$ and maximum out-degree two such that opportunistic greedy search reaches any target $q\in X$ from any starting vertex in $
 O(d\log n)$ 
steps and distance evaluations.
\end{reptheorem}
\begin{proof}
Let $T$ be the Voronoi routing tree and $G$ the corresponding search graph, as in \Cref{def:tree}. Applying \Cref{thm:main} to each $Y \subseteq X$ with $|Y| \ge 2$, \eqref{eq:ordered-balance} holds with $\alpha = \frac{1}{32d}$. Thus, by \Cref{lem:tree-depth}, $T$ has depth at most $h = \lceil \ln n/\alpha \rceil = O(d\log n)$, and by \Cref{lem:routing}, opportunistic greedy search on $G$ reaches any target $q \in X$ from any starting vertex within $2h+1$ steps, each requiring at most two distance evaluations.
\end{proof}

\smallskip

\noindent\textbf{Extensions to Other Metrics.}
\Cref{lem:routing,lem:tree-depth} apply in any metric space. If every SAS consisting of points in $X$ has length at most $L$, then by \Cref{thm:reduction} we can take $\alpha = 1/(4L)$, giving routing in $O(L\log n)$ steps and distance computations. For example, for $X \subseteq \R^d$ under the $\ell_p$ norm with fixed $1 \le p < \infty$, every SAS contained in $X$ has aspect ratio at most $\aspect(X)$. So, by \Cref{thm:bounded-aspect-intro,thm:reduction}, opportunistic greedy search reaches any target $q \in X$ in $O_p\!\left(d^{3q_p/2}\log(2\aspect(X))\log n\right)$ steps (and distance computations) from any starting vertex.

%\section{Why the Routing Model Matters}
%\label{sec:limitations}
% !TEX root = ../main.tex

\section{Exponential SAS Lower Bounds in \texorpdfstring{$\ell_p$}{ℓₚ} Spaces}
\label{sec:lp-extensions}

Our reduction from balanced Voronoi partitions to SASs (\Cref{thm:reduction}) holds in any metric space -- the particular distance function only enters when we bound the length of an SAS. In \Cref{sec:euclidean-sas}, we showed that SASs in $\R^d$ under the $\ell_p$ norm for fixed $1 \le p < \infty$ have length polynomial in $d$ and logarithmic in the aspect ratio (\Cref{thm:bounded-aspect-intro}). In this section, we prove \Cref{thm:negative-non-euclidean-p}, which shows that, without a bound on the aspect ratio, SASs can have length exponential in $d$ for every fixed $p \neq 2$. We do so via two separate theorems: \Cref{thm:negative-large-p}, which covers $1<p<\infty$ with $p \neq 2$, and \Cref{thm:negative-small-p}, which covers $0<p\le 1$. A simple construction, described in \Cref{sec:extensionsIntro}, gives an exponential in $d$ lower bound for $p = \infty$, even with aspect ratio arbitrarily close to one.

%Throughout, we write $\ell_p^d$ for $\R^d$ under the $\ell_p$ norm, and call $u \in \R^d$ a unit vector if $\norm{u}_p = 1$.

\subsection{Exponential Lower Bound for \texorpdfstring{$1<p<\infty$, $p\ne2$}{1 < p < ∞, p ≠ 2}}

We start with our bound for $1 < p < \infty$, with $p \neq 2$. Formally we show:

%Our construction for general $p \neq 2$ follows the same general idea as the $\ell_1$ norm construction described in \Cref{sec:extensionsIntro}. We pick an exponentially sized set of unit vectors $u_1,\ldots,u_n$ such that moving in the direction from $u_i$ to $u_j$ for any $j$ increases the $\ell_p$ norm. I.e., $\norm{u_i-t u_j}_p > \norm{u_i}_p$ for sufficiently small $t$. We note that picking such a set is not possible in Euclidean space -- it would require an exponential number of vectors that all have negative inner product with each other. This is ruled out by \Cref{lem:direction-packing}.  
%We note that our construction has doubly exponential aspect ratio -- as is necessary given \Cref{thm:bounded-aspect-intro}. Formally, we show

%For $u\in\R^d$, define its duality vector $J_p(u) \in \R^d$ coordinatewise by
%\[
% J_p(u)[r]:=|u[r]|^{p-2}u[r],
% \qquad r\in[d],
%\]
%where $J_p(u)[r]=0$ when $u[r]=0$. Note that $J_2(u)=u$.

\begin{theorem}[Exponential SAS Lower Bound for $1<p<\infty$] % jack: this had -- before but it looked like -1<p<\infty
\label{thm:negative-large-p}
For every fixed $1<p<\infty$ with $p\ne2$, there exists an SAS of length $\exp(\Omega_p(d))$ in $\R^d$ under the $\ell_p$ norm.
\end{theorem}
Our proof uses the gradient vector of the $\ell_p$ norm, defined for  $u\in\R^d$ as:

%For $u\in\R^d$, define its duality vector $J_p(u) \in \R^d$ coordinatewise by
\[
 J_p(u)[i]:=|u[i]|^{p-2}u[i],
 \qquad i\in[d],
\]
where $J_p(u)[i]=0$ when $u[i]=0$. Note that $J_2(u)=u$. In  \Cref{app:non-euclidean-lp} we show (\Cref{lem:negative-duality-directions}) that  there are exponentially many unit vectors $u_1,\ldots,u_N$ with $\langle J_p(u_i),u_j \rangle<0$ for all $i \neq j$. I.e., moving a sufficiently small distance from $u_i$ in the direction of $-u_j$ increases the $\ell_p$ norm. We note that such a construction is not possible in Euclidean space -- it would require an exponential number of vectors that all have negative inner product with each other, which is ruled out by \Cref{lem:direction-packing}.  

Like in the $\ell_1$ norm construction  described in \Cref{sec:extensionsIntro}, we scale the vectors with geometrically decreasing values $r_1 > r_2 >\ldots$, and use this first-order increase to argue that the SAS condition (\Cref{def:sas}) holds for base point $k$ and comparison points $i,j$ with $k < i < j$. As for the $\ell_1$ norm, we can check that the condition also holds in the remaining cases when $i < k < j$ and $i < j < k$ simply by using the fact that $\norm{r_i u_i - r_j u_j}_p \approx \norm{r_i u_i}_p$ when $i < j$. Our formal proof follows.

\begin{proof}[Proof of \Cref{thm:negative-large-p}]
Fix $p\in(1,\infty)\setminus\{2\}$. By \Cref{lem:negative-duality-directions}, there are $N=\exp(\Omega_p(d))$ unit vectors $u_1,\ldots,u_N\in \R^d$ and a constant $\gamma_p>0$ such that for all $i \neq k$, 
\begin{equation}
\label{eq:negative-duality}
 \langle J_p(u_k),u_i \rangle\le-\gamma_p.
\end{equation}

Let $f_{k,i}(t) := \norm{u_k-tu_i}_p$. By \Cref{lem:lp-first-order}, $f'_{k,i}(0) =  -\langle J_p(u_k),u_i \rangle \ge \gamma_p$. Also $f_{k,i}(0) = \norm{u_k}_p = 1$. So, there is some $\delta_{k,i}>0$ such that $f_{k,i}(t)\ge 1+\frac{\gamma_p}{2}t$ for all $0<t\le\delta_{k,i}$. Setting $\delta = \min_{k\neq i}\delta_{k,i} > 0$, we have for all $k \neq i$ and $0 < t \le \delta$,
\begin{equation}
\label{eq:uniform-first-order-increase}
 \norm{u_k-tu_i}_p
 \ge 1+\frac{\gamma_p}{2}t.
\end{equation}
Pick scaling factor $\eta > 0$ with $\eta < \min\{1/3,\delta,\gamma_p/2\}$ and let $x_i = \eta^i u_i$. We now will show that $x_1,\ldots,x_N$ form an SAS.
 Fix $i<j$ and a base index $k\ne i$. To satisfy \Cref{def:sas}, we must show
 $\norm{x_k-x_j}_p<\norm{x_k-x_i}_p.$
There are three possible positions of $k$ relative to $i$ and $j$. The first we handle using \eqref{eq:uniform-first-order-increase}. The other two are easy consequences of the triangle inequality.

\smallskip

\noindent\textbf{Case 1: $k<i<j$.}
%Set $t_i:=\eta^{i-k}$ and $t_j:=\eta^{j-k}$. Then $t_i\le\eta\le\delta$ and $t_j\le\eta t_i$. 
By \eqref{eq:uniform-first-order-increase},
\[
 \norm{x_k-x_i}_p
 =\eta^k\norm{u_k-\eta^{i-k} u_i}_p
 \ge\eta^k \left (1+\frac{\eta^{i-k} \gamma_p}{2} \right ).
\]
On the other hand, by triangle inequality, we have $ \norm{x_k-x_j}_p \le \eta^k (1+\eta^{j-k}) \le \eta^k(1+\eta \cdot \eta^{i-k})$. Since we picked $\eta < \gamma_p/2$, this gives that  $\norm{x_k-x_j}_p < \norm{x_k-x_i}_p$ as desired.
%\[
% \norm{x_k-x_j}_p
% =\eta^k\norm{u_k-\eta^{j-k} u_j}_p
% \le\eta^k(1+\eta^{j-k})
% \le\eta^k(1+\eta t_i).
%\]
%Since $\eta<b$, the second quantity is strictly smaller than the first, proving \eqref{eq:sas-goal-lp}.

\smallskip

\noindent\textbf{Case 2: $i<k<j$.}
By triangle inequality we have:
\[
 \norm{x_k-x_j}_p
 \le(1+\eta)\eta^k
 \qquad\text{and}\qquad
 \norm{x_k-x_i}_p
 \ge(\eta^{-1}-1)\eta^k.
\]
Since we picked $\eta<1/3$, we have $1+\eta<\eta^{-1}-1$, giving  $\norm{x_k-x_j}_p<\norm{x_k-x_i}_p$, as desired.

\smallskip

\noindent\textbf{Case 3: $i<j<k$.}
Similarly,
\[
 \norm{x_k-x_j}_p
 \le(1+\eta)\eta^j
 \qquad\text{and}\qquad
 \norm{x_k-x_i}_p
 \ge(\eta^{-1}-\eta)\eta^j.
\]
Again, $\eta<1/3$ implies $1+\eta<\eta^{-1}-\eta$, again giving  $\norm{x_k-x_j}_p<\norm{x_k-x_i}_p.$

These cases exhaust all possibilities (outside the trivial case of $k= j$ when $\norm{x_k-x_j}_p = 0 < \norm{x_k-x_i}_p$). Thus, $(x_1,\ldots,x_N)$ is an SAS of length $\exp(\Omega_p(d))$ as desired.
\end{proof}

We remark that the SAS constructed above has aspect ratio roughly $\eta^{-N}$, which is doubly exponential in $d$. This is required by the upper bound for bounded aspect ratio SASs given in  \Cref{thm:bounded-aspect-intro}.

\subsection{Exponential Lower Bound for \texorpdfstring{$0<p\le1$}{0 < p ≤ 1}}

We now extend our exponential SAS construction to any $0<p \le1$, again following the approach sketched for $\ell_1$ in \Cref{sec:extensionsIntro}. We note that for $0 < p < 1$, $\norm{x-y}_p$ is not in fact a metric -- it may not satisfy triangle inequality. However, $\norm{x-y}_p^p$ is, so we will work with this metric instead. Since the definition of an SAS only involves distance comparisons, and since these are identical under $\norm{x-y}_p$ and $\norm{x-y}_p^p$, this has no impact. Formally we show: % on our conclusions. As in the proof of \Cref{thm:negative-large-p}, we scale our points by geometrically decreasing factors, and verify the SAS condition by considering the same three cases, depending on the position of the base point $k$ relative to the comparison points $i<j$. 

\begin{theorem}[Exponential SAS Lower Bound for $0<p\le1$]
\label{thm:negative-small-p}
For every fixed $0<p\le1$, there exists an SAS of length $\exp(\Omega(d))$ in $\R^d$ under the $\ell_p$ norm.
\end{theorem}

\begin{proof}
Let $m=\lfloor d/8\rfloor$. It is well known that there are $N=\exp(\Omega(d))$ sets $X_1,\ldots,X_N\subseteq[d]$, each of size $m$, with $|X_i\cap X_j|\le m/4$ for all $i\ne j$. This follows, e.g., from the Gilbert--Varshamov bound, and can be proven just by taking a random family of sets \cite{guruswami2010gilbertvarshamov}. Let $r_i=3^{-i/p}$, so that $r_i^p=3^{-i}$, and set $x_i=r_i\cdot\mathbf 1_{X_i}$, where $\mathbf 1_{X_i}\in\{0,1\}^d$ is the indicator vector for $X_i$. For $p=1$, this is exactly the construction sketched in \Cref{sec:extensionsIntro}.

We first bound the distance between any pair of points. Fix $s<t$. Each of the $m-|X_s\cap X_t|$ coordinates in $X_s\setminus X_t$ contributes $r_s^p$ to $\norm{x_s-x_t}_p^p$, each of the $m-|X_s\cap X_t|$ coordinates in $X_t\setminus X_s$ contributes $r_t^p$, and each coordinate in $X_s\cap X_t$ contributes $(r_s-r_t)^p$. Thus,
\begin{align}
 \norm{x_s-x_t}_p^p
 &=(m-|X_s\cap X_t|)\cdot(r_s^p+r_t^p)+|X_s\cap X_t|\cdot(r_s-r_t)^p \nonumber\\
 &\in\left[m\left(r_s^p+\frac{r_t^p}{2}\right),\ m\left(r_s^p+r_t^p\right)\right].
 \label{eq:small-p-distance}
\end{align}
The upper bound uses that $(r_s-r_t)^p \le r_s^p$. The lower bound uses that $(r_s-r_t)^p\ge r_s^p-r_t^p$, and the fact that $|X_s\cap X_t|\le m/4$. 

Now, fix $i<j$ and a base index $k\ne i$. For $(x_1,\ldots,x_N)$ to be an SAS  (\Cref{def:sas}), we must show that $\norm{x_k-x_j}_p^p<\norm{x_k-x_i}_p^p$. As in the proof of \Cref{thm:negative-large-p}, there are three possible positions of $k$ relative to $i$ and $j$. We prove the bound for each separately.
\smallskip

\noindent\textbf{Case 1: $k<i<j$.}
Since $r_j^p\le r_i^p/3<r_i^p/2$,
\[
 \norm{x_k-x_j}_p^p\le m(r_k^p+r_j^p)<m\left(r_k^p+\frac{r_i^p}{2}\right)\le\norm{x_k-x_i}_p^p.
\]

\smallskip

\noindent\textbf{Case 2: $i<k<j$.}
Since $r_j^p\le r_k^p/3$ and $r_k^p\le r_i^p/3$,
\[
 \norm{x_k-x_j}_p^p\le m(r_k^p+r_j^p)\le\frac43\,m r_k^p<m r_i^p\le\norm{x_k-x_i}_p^p.
\]

\smallskip

\noindent\textbf{Case 3: $i<j<k$.}
Since $r_k^p\le r_j^p/3$ and $r_j^p\le r_i^p/3$,
\[
 \norm{x_k-x_j}_p^p\le m(r_j^p+r_k^p)\le\frac43\,m r_j^p<m r_i^p\le\norm{x_k-x_i}_p^p.
\]
Thus, $(x_1,\ldots,x_N)$ is an SAS of length $N=\exp(\Omega(d))$, as desired.
\end{proof}
%As for \Cref{thm:negative-large-p}, the SAS constructed above has aspect ratio roughly $3^{N/p}$, which is doubly exponential in $d$.

%\section{Related Work}
\label{sec:related-work}

\section*{AI Acknowledgement}
A variety of AI tools were used in formulating the proofs  and writing up this manuscript. In particular, the main proof approach of reducing the balanced Voronoi partitioning problem to the self-approaching sequence (SAS) problem, along with the $\poly(d)$ and $O(d)$ upper bounds for SASs in Euclidean space were obtained via conversations with ChatGPT. The  weaker $\exp(d)$ bound given in \Cref{app:exponential} had been previously obtained by the authors, without LLM assistance.
The linear bound in \Cref{thm:Ld} was obtained as an improvement over the initial bound
in \Cref{thm:Ld-poly} using approximately 101 hours of ``thinking time'' with GPT 5.5 and 5.6. The simplified proof that is presented in this work required significant further human work.

Extensions to and lower bounds for other metrics, as well as the lower bound for partitioning with $k$-centers (\Cref{app:kcenter}) were proven largely by the authors, with some LLM assistance. The authors have verified and rewritten all proofs and accept full responsibility for the published contents. 

\medskip

\noindent\textbf{Formalization.} Most of our results have been formalized in Lean
with the help of Claude Opus 5.5. See \url{https://github.com/jacketsj/formalized-sas-and-voronoi-partitions/}. The definitions and statements have been human-checked.
Some of the formalized results give improved constants compared to what we present in the paper. The proofs in the paper focus on simplified arguments
at the cost of  worse constants.

\bibliographystyle{alpha}
\bibliography{refs}

\clearpage

\appendix

\crefalias{section}{appendix} 

\section{Proofs of Basic Geometric Facts}\label{app:geometric}

In this section, we prove some basic facts about Euclidean and $\ell_p$ spaces that we use in our proofs.

The first  states that any set of unit vectors in $d$-dimensional Euclidean space with pairwise inner products bounded by $1/(2d)$ -- i.e., that have pairwise angles close to $90^\circ$ or greater -- has size $O(d)$.

\begin{replemma}{lem:direction-packing}
If unit vectors $u_1,\ldots,u_N\in\R^d$ satisfy
$\langle u_i,u_j \rangle\le 1/(2d)$ for all $i\ne j$, then $N\le4d$.
\end{replemma}

\begin{proof}
Consider the $N \times N$ Gram matrix $G$ with $G_{ij} = \langle u_i,u_j \rangle$. 
Since $u_1,\ldots,u_N$ are unit vectors, the off-diagonal entries of $G$ lie in the range $[-1,1/(2d)]$. For all $x$ in this range, $(x+1)(x-1/(2d)) \le 0$. So $x^2 \le \frac{1}{2d} - (1-\frac{1}{2d}) x$. Applying this bound to the off-diagonal entries we have:
\begin{align}\label{eq:g1}
    \norm{G}_F^2 &= \sum_{i=1}^N G_{ii}^2 + \sum_{i \neq j} G_{ij}^2\nonumber\\
    &\le N + \frac{N(N-1)}{2d} - (1-\frac{1}{2d}) \sum_{i \neq j} G_{ij}.
\end{align}
Now, let $1$ denote the all ones vector of length $N$. Note that since $G$ is positive semidefinite, $\sum_{i,j} G_{ij} = 1^T G1 \ge 0$. This gives:
\begin{align}\label{eq:g2}
    \sum_{i \neq j} G_{ij} = 1^T G 1 - \sum_{i=1}^N G_{ii} \ge - N.
\end{align}
Further, again using that $G$ is positive semidefinite and that $\rank(G)\le d$, 
\begin{align}\label{eq:g3}
\norm{G}_F^2 \ge \frac{\tr(G)^2}{d} = \frac{N^2}{d}.
\end{align}
Plugging \eqref{eq:g2} and \eqref{eq:g3} into \eqref{eq:g1} then gives:
\begin{align*}
    \frac{N^2}{d} &\le N + \frac{N(N-1)}{2d} +(1-\frac{1}{2d})N\\
    &\le 2N + \frac{N^2}{2d}.
\end{align*}
Rearranging  yields the claimed bound.
%
%
% %, and the $d \times d$ covariance matrix $A = \sum_{i=1}^N u_i u_i^T$. Note that $\norm{G}_F^2 = \norm{A}_F^2 \ge \frac{\tr(G)^2}{d} = \frac{N^2}{d}$.
% \[
%  f(t):=(t+1)\left(t-\frac1{2d}\right).
% \]
% For $i\ne j$, we have $-1\le G_{ij}\le1/(2d)$, hence
% $f(G_{ij})\le0$.  Therefore
% \[
%  \sum_{i,j}f(G_{ij})\le Nf(1)
%  =2N\left(1-\frac1{2d}\right).
% \]
% On the other hand,
% \[
%  \sum_{i,j}G_{ij}=\norm{\sum_i u_i}^2\ge0
% \]
% and, writing $A=\sum_i u_i u_i^\top$,
% \[
%  \sum_{i,j}G_{ij}^2=\norm{A}_F^2
%  \ge\frac{(\operatorname{tr}A)^2}{d}=\frac{N^2}{d}.
% \]
% Expanding $f$ gives
% \[
%  \sum_{i,j}f(G_{ij})\ge\frac{N^2}{2d}.
% \]
% Combining the displays yields $N\le4d-2<4d$.
\end{proof}

The second states that there can be at most $O(d)$ nearly equidistant points in $d$-dimensional Euclidean space, with pairwise distances lying between $1$ and $1+O(1/\sqrt{d})$.
\begin{replemma}{lem:rough-equidistant-l2}
There are universal constants $c_1,c_2 >0$ such that if
$x_1,\ldots,x_N\in\R^d$ satisfy
\[
 1\le\norm{x_i-x_j}_2\le1+\frac{c_1}{\sqrt d}\quad \text{ for all }i \neq j
\]
then $N \le c_2 d$.
\end{replemma}

\begin{proof}
Alon~\cite[Proposition~3.1]{alon2009perturbed} proves that there is a universal constant
$c >0$ such that $N=n+1$ points with pairwise distances in
$[1,1+\eps]$, where $1/\sqrt{n} \le\eps\le1/10$, require
\[
 d\ge c\,\frac{\log n}{\eps^2\log(1/\eps)}.
\]
Choose $c_1$ so that $c_1\le 1/10$ and $c_1^2<2c$.  If $n \le d/c_1^2$, then $N = n+1 \le d/c_1^2 + 1$ and the bound holds for $c_2 = 1/c_1^2 + 1$. Otherwise, if $n>d/c_1^2$, take
$\eps=c_1/\sqrt d$.  Then the hypotheses apply, and the bound gives
$\log n \le \frac{c_1^2}{c}\log(\sqrt d/c_1) < 2 \log(\sqrt d/c_1)$. This gives $n \le \frac{d}{c_1^2}$, which contradicts the assumption that $n > \frac{d}{c_1^2}.$
Thus $N =n+1\le d/c_1^2+1 \le c_2 d$ for $c_2 = 1/c_1^2+1$, as required.
\end{proof}

\Cref{lem:rough-equidistant-l2} can be extended to $\ell_p$ spaces as well. Alon and Pudl\'ak \cite{alon2003equilateral} study the setting where the points are all required to have equal $\ell_p$ distances (i.e., where they form an exactly equilateral set). For any $p \ge 1$, they show an upper bound of $O(d^{q_p})$ on the number of such points, where 
\begin{align}\label{eq:qp}
 q_p:=\frac{2p+2}{2p-1}\le4.
\end{align}
It is not hard to extend their approach to point sets that are sufficiently close to equilateral. In particular we show:
\begin{replemma}{lem:rough-equidistant-lp}
For any fixed $1\le p<\infty$, let $q_p = \frac{2p+2}{2p-1} \le 4$. There exist constants $c_{p,1}$ and $c_{p,2}$ (depending on $p$) such that, if
$x_1,\ldots,x_N\in \R^d$ satisfy
\[
 1\le\norm{x_i-x_j}_p\le1+c_{p,1} \cdot d^{-q_p/2} \quad \text{ for all }i \neq j,
\]
then $N \le c_{p,2} \cdot d^{q_p}$.
\end{replemma}

\begin{proof}

Assume for the sake of contradiction that $N > c_{p,2} \cdot d^{q_p}$. Retain just $m = \lfloor c_{p,2} \cdot d^{q_p} \rfloor$ of the points, and consider the matrix $M \in \R^{m \times m}$ with $M_{ij} = 1-\norm{x_i-x_j}_p^p$. $M$ is a near-identity matrix and thus we can argue that $\rank(M) = \Theta(m) = \Theta(d^{q_p})$. We will then show that this is not possible given that the points  lie in $d$-dimensional space. 

\smallskip

\noindent \textbf{Rank lower bound.} 
For all $i \neq j$, we have $1-(1+c_{p,1} \cdot d^{-q_p/2})^p \le M_{ij} \le 0$. We can bound $$1-(1+c_{p,1} \cdot d^{-q_p/2})^p \ge 1-e^{p c_{p,1} \cdot d^{-q_p/2}} \ge -2p c_{p,1} d^{-q_p/2}.$$ The last bound follows as long as we set $c_{p,1}$ small enough so that  $p c_{p,1} d^{-q_p/2} \le 1$, since for $x \le 1$, $1-e^x \ge -2 x$.
So overall, $|M_{ij}| \le 2 p c_{p,1} d^{-q_p/2}$ for all $i \neq j.$ If we set $c_{p,1} < 1/(4p \cdot \sqrt{c_{p,2}})$, we have:
\begin{align}\label{eq:Mmag}
|M_{ij}| < \frac{1}{2\sqrt{c_{p,2}}} d^{-q_p/2} \le \frac{1}{2\sqrt{m}}\quad \text{ for all }i \neq j \quad \text{ and } M_{ii} = 1 \quad \text{for all }i.
\end{align}

% If we set $c_{p,1} < c_{p,2}/(2p)$, the sum of magnitudes of the off-diagonal entries in any row of $M$ is thus bounded by $$m \cdot  2p c_{p,1} d^{-q_p} \le c_{p,2} d^{q_p} \cdot  2 p c_{p,1} d^{-q_p} < 1.$$ Thus, by Gershgorin's circle theorem, all eigenvalues of $M$ are positive and $\rank(M) = m$.

%\smallskip

%\noindent\textbf{Rank upper bound. }
%
%m \times m$ matrix $$
% Fix a large constant $K=K(p)$, put $N=\lfloor Kd^{q_p}\rfloor$, and set
% \[
%  \eps_p(d):=\frac{1}{6p\sqrt{Kd^{q_p}}}.
% \]
% If $m\ge N$, retain only $N$ points.  Define
% \[
%  M_{ij}:=1-\norm{x_i-x_j}_p^p.
% \]
%
Now, let $g$ be an even degree-$k$ polynomial approximating the function $|x|^p$. In particular, satisfying $\max_{x \in [-2,2]} |g(x)-|x|^p| \le \frac{1}{2d \sqrt{m}}$. By Jackson's theorem, to achieve this error bound, as stated in Lemma 4.1 of \cite{alon2003equilateral}\footnote{\cite{alon2003equilateral} has a small bug in that they don't require $g$ to be even, which is needed later for the matrix $\widehat M$ to be symmetric. However, we can always symmetrize $g$ without increasing its degree and only give an improved error bound since $|x|^p$ is itself even.}, it suffices to take 
\begin{align}\label{eq:degreeUpper}
 k \le c_p d^{1/p}m^{1/(2p)},
\end{align}
where $c_p$ is a constant depending on $p$. Consider the matrix $\widehat M \in \R^{m \times m}$ with
\[
 \widehat{M}_{ij} =1-\sum_{r=1}^d g(x_i[r]-x_j[r]).
\]
Note that since $g$ is even, $\widehat M$ is symmetric.
If we set $c_{p,1}$ sufficiently small, our assumption on the pairwise norms ensures that $\norm{x_i-x_j}_p \le 2$ for all $i,j$ and thus $|x_i[r]-x_j[r]| \le 2$ for all $i,j,r$. Thus, by the polynomial approximation bound above, 
$|\widehat M_{ij}-M_{ij}|\le \frac{1}{2\sqrt{m}}$ for all $i,j$.  Combined with \eqref{eq:Mmag} via triangle inequality, we have:
\begin{align}\label{eq:hatM}
|\widehat M_{ij}| < \frac{1}{\sqrt{m}}\quad \text{ for all }i \neq j \quad \text{and} \quad \widehat M_{ii} \in \left [1-\frac{1}{2\sqrt{m}}, 1+\frac{1}{2\sqrt{m}} \right ]\quad \text{ for all } i.
\end{align}
Set $c_{p,2}$ large enough so that $\sqrt{m} \ge 2$. Then together the bounds of \eqref{eq:hatM} give that 
\begin{align}\label{eq:concrete} 
\tr(\widehat M) \ge 3m/4 \quad \text{ and }\quad \norm{\widehat M}_F^2 \le m \cdot (5/4)^2 + m^2\cdot (1/\sqrt{m})^2 \le 3m.
\end{align}
Letting $\lambda_1,\ldots,\lambda_{\rank(\widehat M)}$ be the non-zero eigenvalues of $\widehat M$, by Cauchy-Schwarz,
\begin{align*}
    \tr(\widehat M)^2 = \left (\sum_{i=1}^{\rank(\widehat M)} \lambda_i \right )^2 \le \rank(\widehat M) \cdot \sum_{i=1}^{\rank(\widehat M)} \lambda_i^2 = \rank(\widehat M) \cdot \norm{\widehat M}_F^2 
\end{align*}
By \eqref{eq:concrete}, we thus have 
$\frac{3^2 m^2}{4^2} \le \rank(\widehat M) \cdot 3m$ and so 
\begin{align}\label{eq:rankLower}
\rank(\widehat M) \ge 3m/16.
\end{align}

\smallskip

\noindent \textbf{Rank upper bound.} At the same time, $\widehat M$ is the sum of $d+1$ matrices: the all ones matrix, and $d$ matrices that contain the evaluations of the degree-k polynomials $g(x_i[r]-x_j[r])$ for $r=1,\ldots, d$. Each of these matrices has rank at most $k+1$ and so by \eqref{eq:degreeUpper},
\begin{align}\label{eq:rankUpper}
\rank(\widehat M) \le 1 + d(k+1) \le (2+c_p) d^{1+1/p} m^{1/(2p)}.
\end{align}
Combining \eqref{eq:rankLower} and \eqref{eq:rankUpper} we conclude that 
\begin{align*}
    3m/16 \le (2+c_p) d^{1+1/p} m^{1/(2p)}.
\end{align*}
Rearranging, this gives $m \le c_p' \cdot d^{q_p}$ for some fixed constant $c_p'$. This contradicts the fact that $m = \lfloor c_{p,2}\cdot d^{q_p}\rfloor$ if we set $c_{p,2}$ large enough, completing the proof.
\end{proof}

% NEW---------------------------------------------------------------------------------------------------------------
%-------------------------------------------------------------------------------------------------------------------
\section{Auxiliary Lemmas for \texorpdfstring{$\ell_p$}{ℓₚ} SAS Lower Bounds}
\label{app:non-euclidean-lp}

In this section we prove several facts used in the proofs of
\Cref{thm:negative-large-p,thm:negative-small-p}, which together give our constructions of $\exp(d)$-length SASs in any $\ell_p$ space with $p \neq 2$, stated in  \Cref{thm:negative-non-euclidean-p}.

\begin{lemma}[First-order Expansion of the $\ell_p$ Norm]
\label{lem:lp-first-order}
Fix $1<p<\infty$. For $u\in\R^d$, let $J_p(u)\in\R^d$ be given by $J_p(u)[i]:=|u[i]|^{p-2}u[i]$, where $J_p(u)[i]=0$ when $u[i]=0$. Consider $u,v\in\R^d$ with $\norm{u}_p = \norm{v}_p = 1$. Let $f(t) := \norm{u-tv}_p$. Then $f$ is differentiable at $0$, with $f(0) = 1$ and
\[
 f'(0)=-\langle J_p(u),v \rangle.
\]
\end{lemma}

\begin{proof}
Let
\[
 g(t) = f(t)^p =\norm{u-tv}_p^p
      =\sum_{i=1}^d|u[i]-tv[i]|^p.
\]
Since $p>1$, each term in the sum is differentiable at $t=0$, with derivative $-p|u[i]|^{p-2}u[i]v[i]$.  Therefore,
\[
 g(0)=1
 \qquad\text{and}\qquad
 g'(0)
 =-p\sum_{i=1}^d|u[i]|^{p-2}u[i]v[i]
 =-p\langle J_p(u),v \rangle.
\]
Now $f(t)=g(t)^{1/p}$, and $f(0) = g(0)^{1/p} = 1$. Applying the chain rule we have: 
\[
 f'(0)
 =\frac1p \cdot g(0)^{1/p-1} \cdot g'(0)
 =\frac{1}{p} \cdot 1 \cdot -p\langle J_p(u),v \rangle =  -\langle J_p(u),v \rangle.
\]
\end{proof}

\begin{lemma}[Directions With Negative $\ell_p$ Derivatives]
\label{lem:negative-duality-directions}
Fix $1<p<\infty$ with $p\ne2$.  There are $ N=\exp(\Omega_p(d))$ unit vectors $u_1,\ldots,u_N\in \R^d$ and a constant
$\gamma_p>0$, depending only on $p$, such that, for all $i \neq j$,
\[
 \langle J_p(u_i),u_j \rangle\le-\gamma_p.
\]
\end{lemma}
\begin{proof}
Let $Z$ be the random variable
\begin{align}\label{eq:zVar}
 Z=
 \begin{cases}
  a,&\text{with probability }1/3,\\
  -1,&\text{with probability }2/3,
 \end{cases}
\end{align}
where $a>0$ will be chosen later.  We can compute:
\begin{equation}
\label{eq:two-means}
 \mathbb E[Z]=\frac{a-2}{3}
 \qquad\text{and}\qquad
 \mathbb E[|Z|^{p-2}Z]=\frac{a^{p-1}-2}{3}.
\end{equation}
The first expression vanishes at $a=2$, while the second vanishes at
$a=2^{1/(p-1)}$.  These two values are different because $p\ne2$.
Choose $a$ strictly between them.  The two expectations in
\eqref{eq:two-means} then have opposite signs.

%Let $Z'$ be an independent copy of $Z$.  By independence,
%\begin{align}
% \mu_p
% &:={\mathbb E}\bigl[|Z|^{p-2}ZZ'\bigr]
% \notag\\
% &=\mathbb E[|Z|^{p-2}Z]\,\mathbb E[Z]
% <0.
%\label{eq:negative-mean}
%\end{align}

We  pick random vectors $ v_1,\ldots,v_N\in\{a,-1\}^d$ each with i.i.d. entries  distributed according to \eqref{eq:zVar}.  For $i \neq j$ we have:
\[
 \langle J_p(v_i),v_j \rangle
 =\sum_{r=1}^d
   |v_i[r]|^{p-2}v_i[r]v_j[r].
\]
For independent $Z$ and $Z'$ distributed according to \eqref{eq:zVar}, we have 
$$\mu_p := {\mathbb E}\bigl[|Z|^{p-2}ZZ'\bigr] = \mathbb E[|Z|^{p-2}Z] \cdot \mathbb E[Z] < 0,$$ due to the expectations in \eqref{eq:two-means} having opposite signs. I.e., all terms in the above sum have mean $\mu_p$. Thus, by a standard Bernstein/Hoeffding bound,  there is a constant $c_p>0$, depending only on $p$, such that
\begin{equation}
\label{eq:hoeffding-duality}
 \Pr\left[
  \langle J_p(v_i),v_j \rangle>\frac{\mu_p}{2}d
 \right]
 \le e^{-c_pd}.
\end{equation}
If we take $N= e^{c_pd/3}$, by a union bound over at most $N^2$ pairs of points, the probability that \eqref{eq:hoeffding-duality} occurs for at least one pair is at most $ N^2e^{-c_pd}<1$, for large enough $d$. Thus, there is always a choice of $v_1,\ldots,v_N$ so that for all $i \neq j$
\begin{align}\label{eq:negative-unnormalized-random}
  \langle J_p(v_i),v_j \rangle \le \frac{\mu_p}{2}d.
\end{align}

Normalize each vector by setting $u_i:=\frac{v_i}{\norm{v_i}_p}$. Since we chose $a$ strictly between $2$ and $2^{1/(p-1)}$, both of which are greater than $1$, we have $a > 1$ and so $ \norm{v_i}_p\le a d^{1/p}.$ Further, 
\[
 \langle J_p(u_i),u_j \rangle
 =
 \frac{\langle J_p(v_i),v_j \rangle}
      {\norm{v_i}_p^{p-1}\norm{v_j}_p}.
\]
Thus, by \eqref{eq:negative-unnormalized-random} and the fact that
$\norm{v_i}_p^{p-1}\norm{v_j}_p\le a^pd$, we have
\[
 \langle J_p(u_i),u_j \rangle
 \le \frac{\mu_p}{2a^p}.
\]
This gives the lemma with constant $
 \gamma_p:=\frac{-\mu_p}{2a^p}>0.$
%Finally,
%$N=\lfloor e^{c_pd/3}\rfloor=\exp(\Omega_p(d))$, as required.
\end{proof}

\section{Balanced Partitioning with \texorpdfstring{$k > 2$}{k > 2} Voronoi Cells}\label{app:kcenter}

As discussed, several works \cite{honiden2009balancing,de2015interference,har2016space} ask a question related to ours on balanced Voronoi partitions. They study how many centers must be chosen so that the cells in their Voronoi partition all contain at most an $O(1/k)$ fraction of the input points. Roughly speaking, these papers show that $\tilde O(k)$ centers suffice. However, the constants in their bounds depend exponentially on the dimension $d$ -- see e.g., Theorem 4 of \cite{de2015interference} and Lemma 5.2 of \cite{har2016space}.  \cite{honiden2009balancing} consider Voronoi partitions on bounded degree graphs, showing that on a graph with $n$ nodes and maximum degree $\Delta$, one can always select $O(k \log n)$ nodes so that each cell in their Voronoi partition (with respect to shortest path distance on the graph) contains at most $n\Delta/k$ nodes. One could apply this result to a Euclidean spanner to obtain a balanced partitioning result for Euclidean space. However, such spanners require maximum degree scaling as $\Delta = \exp(d)$  \cite{le2025truly}. %, and thus such an argument would not recover a $\poly(d)$ bound as we do in \Cref{thm:main}.

Given our positive result (\Cref{thm:main}) for the $k = 2$ case, one might hope that these existing works could be improved to depend polynomially on $d$. Perhaps surprisingly, this is not the case: there are point sets in $d$-dimensional Euclidean space such that $\Omega(\exp(d))$ centers must be chosen even for the maximum number of points in any Voronoi cell to be upper bounded by $|X|/4$. We prove this fact below.
\begin{theorem}[Lower Bound for Balanced $k$-center Voronoi Partitions]\label{thm:kwayLower}
    For any $d \ge 1$ there exists a set $X$ of $2^{\Omega(d)}$ points in $\R^d$ such that, for any subset $P \subset X$ with $|P| \le |X|/4$, letting $S(p) = \{x \in X: \norm{x-p}_2 < \min_{q \in P\setminus \{p\}} \norm{x-q}_2\}$ be the set of points assigned to $p$'s Voronoi cell, $$\max_{p \in P} |S(p)| > |X|/4.$$
\end{theorem}
\begin{proof}
Take $d \ge 2$ -- otherwise the claim holds vacuously by letting $X$ contain just a single point. Let $d' = \lfloor d/2 \rfloor$. It is well known that there exist sets of $2^{\Omega(d')} = 2^{\Omega(d)} $ unit vectors in $\R^{d'}$ with all pairwise inner products bounded by $1/4$ and thus all pairwise distances at least $\sqrt{3/2}$. This follows e.g., from the Gilbert--Varshamov bound, and can be proven just by taking a random set of unit vectors \cite{guruswami2010gilbertvarshamov}. Let $U = \{u_1,u_2,\ldots,u_{|U|}\}$ be such a set. Pick scalars $r_1 > r_2 > \ldots > r_{|U|} > 0$ and consider the set of points $U' = \{ r_1 u_1,\ldots,r_{|U|} u_{|U|}\}$. We will pick these scalings so that they are arbitrarily close to $1$ -- i.e., so that the vectors in $U'$ are nearly unit vectors.

Let $X_1$ be the set of points in $\R^d$ given by taking the points of $U'$ as the first $d'$ coordinates, and leaving all other coordinates equal to $0$. Similarly, let $X_2$ be the set of points  in $\R^d$ given by taking the points of $U'$ as the $d'+1,\ldots, 2d'$ coordinates, scaling them by a factor of $1/2$, and leaving all other coordinates equal to $0$. Let $X = X_1 \cup X_2$.

By construction, any $y \in X_1$ and $z \in X_2$ are orthogonal to each other. Thus, we can ensure $\norm{y-z}_2 = \sqrt{\norm{y}_2^2 + \norm{z}_2^2}$ is arbitrarily close to $\sqrt{5/4}$ by taking our scalings $r_1 > r_2 > \ldots > r_{|U|}$  arbitrarily close to $1$. At the same time, for any $y,y' \in X_1$, again by taking these scalings arbitrarily close to $1$, we have $\norm{y-y'}_2$ arbitrarily close to being lower bounded by $\sqrt{3/2} > \sqrt{5/4}$. In other words, all points in $X_2$ are closer to all points in $X_1$ than the points in $X_1$ are to each other.

Now, consider two cases. \textbf{Case 1:} The set of centers $P$ contains no points in $X_2$. In this case, $|S(p)| \ge |X_2| = |X|/2$ for the point $p \in P$ (and thus also $p \in X_1$) with smallest scaling, since $p$ will be the closest center to all of the (orthogonal) points in $X_2$. \textbf{Case 2:} The set of centers $P$ contains a point in $X_2$. In this case, letting $p \in P \cap X_2$ be the center in $X_2$ with the smallest scaling, $|S(p)| \ge |X_1| - |X_1 \cap P|$ -- i.e., $p \in X_2$ will be the closest center to all points in $X_1$, except those that are themselves chosen as centers. Since $|X_1 \cap P| < |P| \le |X|/4$ by assumption, this gives that $|S(p)| > |X|/4$.
So in both cases, $\max_{p \in P} |S(p)| > |X|/4$ as required.
\end{proof}

% !TEX root = ../main.tex

\section{Exponential Bounds via Navigable Graphs}
\label{app:exponential}

In this section, we show that classic constructions of navigable graphs in $d$-dimensional Euclidean space with degree scaling exponentially in $d$ \cite{AryaMount:1993} can be used to show the existence of $\alpha$-balanced Voronoi partitions with $\alpha = 1/2^{O(d)}$. Our main result on balanced partitions, \Cref{thm:main}, improves this bound to $\Omega(1/d)$.

Recall that a directed graph on a point set $X$ is \emph{navigable} if greedy search initialized at any starting point successfully navigates to any target $q \in X$. Equivalently, for every pair of distinct points $x,q \in X$, there is an out-neighbor $y$ of $x$ with $D(y,q) < D(x,q)$. Arya and Mount~\cite[Section~3 and Lemma~3.1]{AryaMount:1993} show the following bound for navigable graphs in Euclidean space: 

\begin{lemma}[$\exp(d)$-degree Navigable Graphs]
\label{lem:exponential-navigable}
Every set $X \subseteq \R^d$ admits a navigable graph under the Euclidean distance with maximum out-degree at most $3^d$.
\end{lemma}
We include a proof below for completeness and to remove an assumption of \cite{AryaMount:1993} that the points lie in general position.
\begin{proof}
For each $x \in X$, construct its out-neighborhood as follows. Initially, all points in $X \setminus \{x\}$ are candidates. Repeatedly choose a candidate $y$ minimizing $\norm{x-y}_2$, add the edge $x \to y$, and remove $y$ together with every remaining candidate $z$ satisfying $\norm{z-y}_2 < \norm{z-x}_2.$ When choosing $y$, break any ties  arbitrarily. 
Every target $q \in X \setminus \{x\}$ is either selected as an out-neighbor of $x$ or removed by an out-neighbor strictly closer to $q$ than $x$ is. Thus, the resulting graph is navigable.

To bound the out-degree, consider two selected neighbors $y,z$ of $x$, with $y$ selected first. Write $\frac{y-x}{\norm{y-x}_2} = u$ and $\frac{z-x}{\norm{z-x}_2} =  v$. Since $z$ was not removed when $y$ was selected, $\norm{(z-x)-(y-x)}_2 = \norm{z-y}_2 \ge \norm{z-x}_2$. Squaring gives
\[
 \norm{z-x}_2^2+\norm{y-x}_2^2-2 \cdot \norm{z-x}_2 \cdot \norm{y-x}_2 \cdot \langle u,v \rangle \ge \norm{z-x}_2^2.
\]
Rearranging, we have  
\begin{align*}
 \langle u,v \rangle \le \frac{\norm{y-x}_2}{2 \norm{z-x}_2} \le \frac12.
 \end{align*}
Consequently, the unit directions of any two out-neighbors have distance at least $1$. The open balls of radius $1/2$ centered at these directions are pairwise disjoint and contained in the ball of radius $3/2$ centered at the origin. Comparing volumes bounds their number by $3^d$, as required.
\end{proof}

The next lemma is a generic conversion from a navigable graph with bounded out-degree to a balanced Voronoi partition.

\begin{lemma}[Bounded Degree Navigability Implies Balanced Partitioning]
\label{lem:navigable-balance}
Let $X$ be a set of $n \ge 2$ points in a metric space $(\mathcal X,D)$. If $X$ admits a navigable graph with maximum out-degree at most $M$, where $M \ge 2$, then $X$ admits an $\alpha$-balanced Voronoi partition for $\alpha = 1/(2M)$.
\end{lemma}
\begin{proof}
Recall from \Cref{sec:reduction} that $S(a,b)$ is the set of points in $X$ closer to  $a$ than $b$, with ties broken by id.
Suppose for the sake of contradiction that $X$ has no $\frac{1}{2M}$-balanced Voronoi partition. Fix any $x \in X$. By navigability, every point in $X \setminus \{x\}$ belongs to $S(y,x)$ for some out-neighbor $y$ of $x$. Since there are at most $M$ out-neighbors, at least one such $y$ satisfies
\[
 |S(y,x)| \ge \frac{n-1}{M} \ge \frac{n}{2M}.
\]
Our assumption that there is no $\frac{1}{2M}$-balanced Voronoi partition thus  implies that $|S(x,y)| < \frac{n}{2M}$.

Choose one such out-neighbor for each $x$. Starting from any point and repeatedly following the chosen edge must eventually produce a directed cycle. Denote its vertices by $a_1,\ldots,a_m$, and set $a_{m+1}=a_1$. Then we have
\begin{equation}
\label{eq:cycle-small-cell}
 |S(a_i,a_{i+1})| < \frac{n}{2M}
 \qquad\text{for all }1 \le i \le m.
\end{equation}
We will show that some point on the cycle gives a balanced partition together with $a_1$.
For $2 \le i \le m$, let
\[
 O_i := \frac{|S(a_i,a_1)|}{n}.
\]
Since $S(a_1,a_2) \cup S(a_2,a_1) = X$, \eqref{eq:cycle-small-cell} gives $O_2 > 1-\frac{1}{2M}$. The final edge of the cycle gives $O_m < \frac{1}{2M}$. Further, for $2 \le i < m$,
\[
 S(a_i,a_1) \setminus S(a_i,a_{i+1})
 \subseteq S(a_{i+1},a_1).
\]
Indeed, a point $x$ in the set on the left is closer to $a_{i+1}$ than $a_i$ and to $a_i$ than $a_1$, and thus is closer to $a_{i+1}$ than $a_1$. Thus, by \eqref{eq:cycle-small-cell},
\[
 O_{i+1} \ge O_i - \frac{|S(a_i,a_{i+1})|}{n}
 > O_i - \frac{1}{2M}.
\]
In other words, the $O_i$ cannot decrease by more than $\frac{1}{2M}$ in any step. Let $k$ be the first index for which $O_k \le 1/2$. Such an index exists since $O_m < \frac{1}{2M} \le 1/4$ (by our assumption that $M \ge 2$), and $k \ge 3$ since $O_2 > 1-\frac{1}{2M} \ge \frac{3}{4}$. The preceding bound implies
\[
 \frac{|S(a_k,a_1)|}{n} = O_k > O_{k-1}-\frac{1}{2M} > \frac12-\frac{1}{2M} \ge \frac{1}{2M}.
\]
Also, $S(a_1,a_k) \cup S(a_k,a_1) = X$, so
\[
 \frac{|S(a_1,a_k)|}{n} \ge 1-O_k \ge \frac12 \ge \frac{1}{2M}.
\]
Hence $a_1,a_k$ give a $\frac{1}{2M}$-balanced Voronoi partition (\Cref{def:balancedPartition}), contradicting our assumption that no such partition existed. %The argument uses the weak inequalities in \Cref{def:balancedPartition}, so no general-position assumption is needed.
\end{proof}

Combining \Cref{lem:exponential-navigable,lem:navigable-balance} finally gives:
\begin{theorem}[Exponential Balanced Partition Bound  via Navigable Graphs]
\label{thm:exponential-bound}
Any set of points $X$ in $\R^d$ admits an $\alpha$-balanced Voronoi partition under the Euclidean distance for $\alpha = 1/(2 \cdot 3^d)$.
\end{theorem}
%\begin{proof}
%The partitioning bound follows from \Cref{lem:exponential-navigable,lem:navigable-balance} with $K = 3^d$. Now let $(x_1,\ldots,x_\ell)$ be any SAS in $\R^d$ with $\ell \ge 2$, and apply the bound to $X = \{x_1,\ldots,x_\ell\}$. For every $i < j$, the SAS condition gives
%\[
% \norm{x_k-x_j}_2 < \norm{x_k-x_i}_2
% \qquad\text{for all }k \ne i.
%\]
%Thus $S(x_i,x_j) = \{x_i\}$, and every pair of sites has a Voronoi cell containing exactly one point. The partitioning bound therefore implies $1 \ge \ell/(2 \cdot 3^d)$, or $\ell \le 2 \cdot 3^d$, as required.
%\end{proof}

%\input{Sections/linear_sas} % moved to Section 3.3

\end{document}